\documentclass[11pt,a4paper]{article}

\usepackage{amsmath}
\usepackage[psamsfonts]{amsfonts}
\usepackage{theorem}
\usepackage{version}
\usepackage{graphicx}
\usepackage{pstricks,pst-node,pst-text,pst-3d}
\usepackage{subfigure}

\newtheorem{Lem}{\sc Lemma}
\newtheorem{Prop}{\sc Proposition}

\newtheorem{theor}{\sc Theorem}

\usepackage{hyperref}
\hypersetup{pdfpagemode=FullScreen,  
colorlinks=true,
citecolor=blue,
linkcolor=blue,
urlcolor=blue
}

\def\Black{} 

\newcommand{\n}{^{(n)}}

\newcommand{\ptwoone}{p}

\newcommand{\cqfd}{\hfill $\square$}

\newcommand{\R}{\mathbb R}

\newcommand{\N}{\mathbb N}

\newcommand{\Xb}{\mathbf{X}}

\newcommand{\Zb}{\mathbf{Z}}

\newcommand{\ub}{\ensuremath{\mathbf{u}}}

\newcommand{\xb}{\ensuremath{\mathbf{x}}}

\newcommand{\yb}{\ensuremath{\mathbf{y}}}
\newcommand{\zb}{\ensuremath{\mathbf{z}}}
\newcommand{\Ab}{\ensuremath{\mathbf{A}}}

\newcommand{\SSbb}{\ensuremath{\mathbf{S}}}

\newcommand{\Ub}{\ensuremath{\mathbf{U}}}

\newcommand{\Ob}{\ensuremath{\mathbf{O}}}
\newcommand{\zerob}{\ensuremath{\mathbf{0}}}
\newcommand{\Yb}{\ensuremath{\mathbf{Y}}}

\newcommand{\mub}{\ensuremath{\pmb\mu}}

\newcommand{\thetab}{{\pmb \theta}}

\newcommand{\Sigmab}{{\pmb \Sigma}}

\newcommand{\deltab}{{\pmb \delta}}

\newcommand{\ny}{n\rightarrow\infty}

\begin{document}


\title{Depth-based Runs Tests for Bivariate Central Symmetry}
\date{}
  \author{Rainer Dyckerhoff\footnote{Seminar f\"{u}r Wirtschafts- und Sozialstatistik, Universit\"{a}t zu K\"{o}ln, Meister-Ekkehart-Str. 9, 50923 K\"{o}ln, Germany}, \; Christophe Ley\footnote{ECARES and D\'{e}partement de Math\'{e}matique, Universit\'{e} libre de Bruxelles, CP 210, Bld. du Triomphe, B-1050, Brussels, Belgium} \; 
and Davy Paindaveine\footnote{ECARES and D\'{e}partement de Math\'{e}matique, Universit\'{e} libre de Bruxelles, CP 114/04, Av. \mbox{F.D.} Roosevelt, 50, B-1050, Brussels, Belgium}} 

\maketitle
 
\begin{abstract}
McWilliams (1990) introduced a nonparametric procedure based on runs for the problem of testing univariate symmetry about the origin (equivalently, about an arbitrary specified center). His procedure first reorders the observations according to their absolute values, then rejects the null when the number of runs in the resulting series of signs  is too small. This test is universally consistent and enjoys nice robustness properties, but is unfortunately limited to the univariate setup. In this paper, we extend McWilliams' procedure into tests of bivariate central symmetry. The proposed tests first reorder the observations according to their statistical depth in a symmetrized version of the sample, then reject the null when an original  concept of simplicial runs is too small. Our tests are affine-invariant and have good robustness properties. In particular, they do not require any finite moment assumption. We derive their limiting null distribution, which establishes their asymptotic distribution-freeness. We study their finite-sample properties through Monte Carlo experiments, and conclude with some final comments.
\end{abstract}


\section{Introduction}

Symmetry is an essential and ubiquitous notion in statistics, and particularly so in multivariate nonparametric statistics. In bivariate location problems, for instance, \emph{all} nonparametric tests do require some concept of symmetry, which may be either {\em spherical symmetry} ($\Xb-\mub\stackrel{d}{=}\Ob(\Xb-\mub)$ for all orthogonal $2\times 2$ matrices~$\Ob$), {\em elliptical symmetry} ($\Xb\stackrel{d}{=}\Ab\Zb+\mub$, where~$\Zb$ is spherically symmetric about the origin of~$\R^2$ and $\Ab$ is an arbitrary $2\times 2$ matrix), or {\em central symmetry} ($\Xb-\mub\stackrel{d}{=}-(\Xb-\mub)$). Consequently, it is crucial to dispose of good tests for---spherical, elliptical and central---symmetry, which can serve as important preliminaries before applying the corresponding location tests. 

In this paper, we focus on testing for bivariate central symmetry, which, in practice, may be more important than testing for spherical or elliptical symmetry---non-rejection of the null of central symmetry indeed justifies resorting to location tests that require a weaker symmetry assumption (central symmetry), hence are more robust than their spherical or elliptical counterparts. Unfortunately, there are much less tests for central symmetry than for spherical or elliptical symmetry; we refer to \cite{Ser2006B} for an extensive review on multivariate symmetry concepts and multivariate symmetry testing.

Now, the tests for central symmetry available in the literature---e.g., those from \cite{Blo1989}, \cite{GhoRuy1992}, \cite{Heaetal1995}, \cite{NeuZhu1998} or \cite{Henetal2003}---are hardly satisfactory : they either do not meet fundamental properties such as, e.g., affine-invariance or (asymptotic) distribution-freeness under the null, or do require stringent moment assumptions. As an illustration, among the precited tests, only the procedures from \cite{Henetal2003} are affine-invariant, but unfortunately they need finite fourth-order moments and are not (not even asymptotically) distribution-free.

We intend to improve on that by proposing tests for bivariate central symmetry that are affine-invariant,  asymptotically distribution-free under the null, and that do not require any moment assumption---more generally, that exhibit good robustness properties. We will achieve this by extending to the bivariate setup the celebrated \cite{McW1990} test. 
This test, which aims at testing the null that the common distribution of the (i.i.d.) observations~$X_i$, $i=1,\ldots,n$, is symmetric about the origin, say\footnote{Also in the bivariate case, we will throughout test for central symmetry about the origin, which is clearly without any loss of generality since testing symmetry about any other fixed value~$\mub_0\in\R^2$ would just be achieved by applying the proposed origin-based tests to the centered observations $\Xb_i-\mub_0$, $i=1,\ldots,n$.},
proceeds in two steps:
\begin{enumerate}
\item[(i)] the sample is reordered  into  $X_{A_1},X_{A_2},\ldots,X_{A_n}$ according to $|X_{A_1}|\leq |X_{A_2}|\leq \ldots\leq |X_{A_n}|$, an ordering that is uniquely defined with probability one under absolute continuity (the $A_i$'s are sometimes called the \emph{anti-ranks} of the~$|X_i|$'s); \label{stepone}
\item[(ii)] the number of runs---$R^{(n)}$, say---in the sequence of signs~${\rm Sign}(X_{A_1}),\ldots,{\rm Sign}(X_{A_n})$ is recorded (the term \emph{run} refers to a maximal sequence of consecutive ones or minus ones), and the null hypothesis is rejected for small values of~$R^{(n)}$.
\end{enumerate}

What makes it natural to try and turn this test into a test for bivariate central symmetry is its many good properties. The test enjoys strong invariance properties (see Section~\ref{defsec}), yielding exact distribution-freeness under the null. It is fairly robust to outliers and does not require any moment assumption.
 More importantly, it has been shown to be consistent against any asymmetric alternative associated with an \mbox{a.e.} continuous density~$f$ (\cite{Hen1993}) and to be very competitive compared to other universally consistent tests (\cite{McW1990}). Finally, it is extremely simple to implement~: the test statistic~$R\n$ is computationally simple, and for large sample sizes, the test can be based on the (null) asymptotic standard normal distribution of~$n^{-1/2}(2R^{(n)}-n)$; see \cite{CohMen1988}.



Extending McWilliams' test to the multidimensional setup requires defining appropriate multivariate versions of Steps~(i)-(ii) above. Compared to the spherically symmetric construction from \cite{Mar1999} (that results into a test for spherical symmetry; see Section~\ref{runssimustest} below), our proposal is of a more nonparametric nature. More specifically, we propose the following bivariate extensions of Steps~(i)-(ii): 
\begin{itemize} 
\item[(i)] In the univariate case, the random permutation does not require a genuine distance from the null symmetry center, but rather only a center-outward ordering of the observations. In the bivariate setup, it therefore seems natural to order observations according to some \emph{statistical depth function} (\cite{ZuoSer2000A}), which is actually providing such a two-dimensional center-outward ordering. 

\item[(ii)] In McWilliams' runs test statistic, a new run is obtained exactly when ${\rm Sign}(X_{A_i})\neq {\rm Sign}(X_{A_{i-1}})$, or equivalently, when the origin is contained in the interval with end points~$X_{A_i}$ and $X_{A_{i-1}}$---an interval that can be seen as the {\em simplex} with vertices~$X_{A_i}$ and $X_{A_{i-1}}$. In the bivariate setup, this suggests defining a runs statistic as the number of simplices with vertices~$\Xb_{A_{i}},\Xb_{A_{i-1}},\Xb_{A_{i-2}}$ that contain the origin. 
\end{itemize}

As we show below, the resulting bivariate McWilliams tests, as desired, are tests of \emph{central} symmetry, and enjoy many nice properties of their univariate antecedent.

The paper is organized as follows. 
Section~\ref{ourrunssec} describes the proposed test statistics~: first, the concept of statistical depth functions is shortly discussed (Section~\ref{depthsec}) and the simplicial bivariate runs are defined (Section~\ref{multrunssec}); then the proposed test statistics are provided and their invariance properties are studied
(Section~\ref{defsec}). 
 In Section~\ref{runsdist}, the  null  asymptotic distribution of our tests is derived, which establishes in particular their asymptotic distribution-freeness. 
Section~\ref{runssimus} is dedicated to Monte Carlo experiments~: several competitors are briefly described (Section~\ref{runssimustest}) before the  Monte Carlo study is conducted (Section~\ref{runssimusas}). 
 Section~\ref{runsFC} provides some final comments. Eventually, the Appendix collects technical proofs.

\section{The proposed tests}\label{ourrunssec}

Consider the null hypothesis~$\mathcal{H}_0^{\rm centr}$ under which the bivariate observations~$\Xb_i$, $i=1,\ldots,n$, are mutually independent and admit a common pdf~$f$ (with respect to the Lebesgue measure on~$\R^2$) that satisfies~$f(\xb)=f(-\xb)$ almost everywhere in~$\xb\in\R^2$ (central symmetry about the origin). 
As announced in the Introduction, we propose tests for bivariate central symmetry that rely on (i) a random permutation of the observations determined by some statistical depth function, and on (ii) an original concept of bivariate runs. Sections~\ref{depthsec} and~\ref{multrunssec} respectively focus on these two aspects.

\subsection{Statistical depth functions}
\label{depthsec}

Informally, a $k$-variate statistical depth function~$D(\cdot,{\rm P}):\R^k \to [0,1]$  provides, for each~$\xb\in\R^k$, a measure~$D(\xb,{\rm P})$ of how central~$\xb$ is relative to the probability measure~${\rm P}$ over~$\R^k$ (the larger~$D(\xb,{\rm P})$ the more central~$\xb$). According to \cite{ZuoSer2000A}, a statistical depth function should satisfy the following four properties:
\begin{enumerate}
\item[P1]
\emph{affine-invariance}: for any invertible $k\times k$ matrix~$\Ab$ and any $\mathbf{b}\in\R^k$, $D(\Ab\xb+\mathbf{b},{\rm P}_{\Ab,\mathbf{b}})=D(\xb,{\rm P})$ where,  for any $k$-dimensional Borel set~$B$,~${\rm P}_{\Ab,\mathbf{b}}[B]:={\rm P}[\Ab^{-1}(B-\mathbf{b})]$; 
\item[P2] \emph{maximality at center}: if ${\rm P}$ is centrally symmetric\footnote{\cite{ZuoSer2000A} also consider P2 for weaker concepts of symmetry, namely angular and halfspace symmetry, but, for our purposes, we may restrict to central symmetry in the sequel.} about~$\xb_0\in\R^k$, then $D(\xb_0,{\rm P})\geq D(\xb,{\rm P})$ for any $\xb\in\R^k$;
\item[P3] \label{tutuu} \emph{monotonicity relative to any deepest point}: if~$D(\xb_0,{\rm P})\geq D(\xb,{\rm P})$ for any $\xb\in\R^k$, then $\lambda \mapsto D(\xb_0+\lambda (\xb-\xb_0),{\rm P})$ is monotone non-increasing over~$[0,\infty)$;
\item[P4] \emph{vanishing at infinity}: as $\|\xb\|\to\infty$, $D(\xb,{\rm P})\to 0$. 
\end{enumerate}
The properties P1-P3 directly entail that statistical depth functions induce an affine-invariant \emph{center-outward ordering} of points in~$\R^k$, where the (depth) center---\mbox{i.e.},  the deepest point---coincides, for symmetric distributions, with the symmetry center.

Classical examples of statistical depths include
\begin{enumerate}
\item The \cite{Tuk1975} {\em halfspace depth} 
$
D^{H}(\xb,{\rm P})
=
\inf_{H\in\mathcal{H}_\xb} {\rm P}[H]$,
where $\mathcal{H}_\xb$ stands for the collection of closed halfspaces in~$\R^k$ with~$\xb$ on their boundary hyperplane;
\item The  \cite{Liu1990} {\em simplicial depth} 
$
D^S(\xb,{\rm P})={\rm P}[\xb\in S(\Xb_1,\Xb_2,\ldots,\Xb_{k+1})],
$
where the $\Xb_i$'s are \mbox{i.i.d.} with common distribution~${\rm P}$ and $S(\xb_1,\xb_2,\ldots,\xb_{k+1})$ stands for the closed simplex with vertices $\xb_1,\xb_2,\ldots,\xb_{k+1}$ in~$\R^k$;
\item The  {\em simplicial volume depth} (sometimes also referred to as {\em Oja depth} in the literature)
$
D^{SV}(\xb,{\rm P})=
1/
\big[
1+
{\rm E}_{\rm P}[
m_k(S(\xb,\Xb_1,\Xb_2,\ldots,\Xb_{k}))
]
\big]
,
$
where the $\Xb_i$'s are \mbox{i.i.d.} with common distribution~${\rm P}$ and $m_k$ denotes the Lebesgue measure in~$\R^k$. This depth does not satisfy~P1; however, if ${\pmb\Sigma}_{\rm P}$ is some affine-equivariant scatter matrix functional (in the sense that~${\pmb\Sigma}_{{\rm P}_{\Ab,\mathbf{b}}}=\Ab{\pmb\Sigma}_{{\rm P}}\Ab'$ for any invertible $k\times k$ matrix~$\Ab$ and any $k$-vector~$\mathbf{b}$), then the modified simplicial volume depth~$D^{SV}_{\rm mod}=
1/
\big[
1+
(\det {\pmb\Sigma}_{\rm P})^{-1/2}
{\rm E}_{\rm P}[
m_k(S(\xb,\Xb_1,\Xb_2,\ldots,\Xb_{k}))
]
\big]
$
satisfies~P1. 
\end{enumerate}

The corresponding deepest points $\thetab_{D^H}$, $\thetab_{D^S}$ and $\thetab_{D^{SV}} (=\thetab_{D^{SV}_{\rm mod}})$  are called the Tukey median, the simplicial median and the \cite{Oja1983} median, respectively. In the univariate case, they all reduce to the univariate median, which justifies the terminology. 

Of course, whenever $k$-variate observations~$\Xb_i$, $i=1,\ldots, n$, are available, sample depth functions are simply obtained as~$\xb\mapsto D(\xb,{\rm P}\n$), where~${\rm P}\n$ denotes the corresponding empirical distribution.
As their population counterparts, sample depth functions  are providing a center-outward ordering with respect to the corresponding deepest point or multivariate sample median, $\thetab\n_D$ say. 

In the univariate McWilliams' test statistic (see Page~\pageref{stepone}), however, observations are permuted according to a center-outward ordering with respect to the null symmetry center---namely the origin of the real line---and not with respect to the median. To properly extend the McWilliams test to the bivariate setup in the sequel, we therefore replace~$D(\cdot, {\rm P}\n)$ with~$D(\cdot, {\rm P}_{\rm sym}\n)$, where~${\rm P}_{\rm sym}\n$ denotes the empirical distribution of the symmetrized sample~$(\pm\Xb_1,\ldots,\pm \Xb_n)$ of size~$2n$. Clearly, it follows from~P2 that the deepest point then is the origin of~$\R^2$, hence that the resulting center-outward ordering is indeed relative to the null symmetry center. 

In the univariate case, the three depth functions above, in their symmetrized versions, will make $x$ deeper than $y$ iff $|x|<|y|$. Therefore, the three resulting center-outward orderings, unlike the statistical depth functions themselves, do strictly agree, and lead to the same ordering as in Step~(i) of the McWilliams procedure.

\subsection{Simplicial runs}
\label{multrunssec}

As mentioned in the Introduction, the univariate \cite{McW1990} test statistic is based on the number of runs in some given sequence. This number of runs, in an ordered real sequence~$x_1,\ldots,x_n$, can be written as
$
1+\sum_{i=2}^n
\,
\mathbb{I}_{[{\rm Sign}(x_i)\neq {\rm Sign}(x_{i-1})]}.
$
Our bivariate extension is motivated by the fact that the same runs statistic can also be expressed as 
\Black
$
1+\sum_{i=2}^n 
\,
\mathbb{I}_{
[0 \in S(x_i,x_{i-1})]}
,
$
where~$S(x,y)=[\min(x,y),\max(x,y)]$ stands for the simplex with vertices~$x,y\in\R$, that is, for the convex hull of those two points on the real line. 

For a sequence of bivariate vectors $\xb_1,\ldots,\xb_n$, it is then natural to define the number of (simplicial) runs as
\begin{equation}\label{ffg}
1+\sum_{i=3}^n
\,
\mathbb{I}_{\displaystyle[\zerob\in S(\xb_i,\xb_{i-1},\xb_{i-2})]},
\end{equation}
where $S(\xb,\yb,\zb)$ still denotes the closed simplex with vertices $\xb,\yb,\zb\in\R^2$. The connection between~(\ref{ffg}) and the bivariate simplicial depth (of the origin of~$\R^2$) is obvious; see \cite{Liu1990} or Section~\ref{depthsec}. Clearly, the ordering of the $\xb_i$'s explains that~(\ref{ffg}) avoids the $U$-statistic structure that characterizes the sample simplicial depth.  


\subsection{The proposed test statistics} 
\label{defsec}


Let~$\Xb_i$, $i=1,\ldots,n$, be bivariate observations and let $D$ be a statistical depth function on~$\R^2$. Sections~\ref{depthsec} and~\ref{multrunssec} lead to extending the univariate \cite{McW1990} test statistic into 
$$
R^{(n)}_{D}
=
1+\sum_{i=3}^n
\,
\mathbb{I}_{\displaystyle [\zerob\in S(\Xb_{A_i},\Xb_{A_{i-1}},\Xb_{A_{i-2}})]},
$$
where the reordered observations~$\Xb_{A_1},\ldots, \Xb_{A_n}$ are defined through 
\begin{equation} \label{dep}
D(\Xb_{A_1},{\rm P}^{(n)}_{\rm sym})
\geq 
D(\Xb_{A_2},{\rm P}^{(n)}_{\rm sym})
\geq 
\ldots
\geq 
D(\Xb_{A_n},{\rm P}^{(n)}_{\rm sym});
\end{equation}
as in Section~\ref{depthsec}, ${\rm P}^{(n)}_{\rm sym}$ stands for the empirical distribution of the symmetrized sample~$(\pm\Xb_1,\ldots,\pm\Xb_n)$. If ties occur in~(\ref{dep}), we impose that each block of undefined anti-ranks~$A_{j+1},\ldots,A_{j+r}$ forms a monotone increasing sequence (we avoid breaking the ties randomly as this would possibly affect affine-invariance of~$R^{(n)}_{D}$; see Proposition~\ref{runsaffinv} below).
Parallel to the univariate case, the resulting bivariate test for central symmetry---$\phi_D^{(n)}$, say---rejects~$\mathcal{H}_0^{\rm centr}$ for small values of the number of simplicial runs $R^{(n)}_{D}$. Critical values will be derived in Section~\ref{runsdist} below.

%
As mentioned in the Introduction, the univariate McWilliams statistic~$R\n$ enjoys strong invariance properties. It is indeed straightforward to check that~$R\n$ is invariant under any transformation of the form
\begin{eqnarray}
g_h
: \,
\R\times\ldots\times \R \ \,
& \to &
\ \R\times\ldots\times \R
\label{univransf}
\\
(x_1,\ldots,x_n) 
\hspace{2mm}
&\mapsto&
\, (h(x_1),\ldots,h(x_n)), 
\nonumber
\end{eqnarray}
where~$h:\R\to\R$ is an odd, continuous, and monotone increasing function satisfying~$h(+\infty)=+\infty$. All such transformations form a group~$\mathcal{G},\circ$ that happens to generate the null hypothesis of symmetry about zero. The exact distribution-freeness of~$R\n$ under the null is a direct corollary of this invariance under a generating group.  
 
One might wonder whether our bivariate statistics~$R\n_D$ are similarly invariant under a group of transformations that generates the null~$\mathcal{H}_0^{\rm centr}$ of central symmetry about the origin. Unfortunately, the answer is negative. Actually, for each of the three depth functions~$D^H$, $D^S$, and~$D^{SV}/D^{SV}_{\rm mod}$ introduced in Section~\ref{depthsec}, it can be checked that $R\n_D$ fails to be invariant under the group of radial transformations
\begin{eqnarray*}
g_h
: 
\R^2\times\ldots\times \R^2
& \to &
\R^2\times\ldots\times \R^2
\\
(\mathbf{x}_1,\ldots,\mathbf{x}_n) 
\hspace{2mm}
&\mapsto&
\bigg(
h^{\|\cdot\|}(\|\mathbf{x}_1\|) \,\frac{\mathbf{x}_1}{\|\mathbf{x}_1\|}
,\ldots,
h^{\|\cdot\|}(\|\mathbf{x}_n\|) \,\frac{\mathbf{x}_n}{\|\mathbf{x}_n\|}
\bigg), 
\end{eqnarray*}
where~$h^{\|\cdot\|}:\R^+\to\R^+$ is continuous, monotone increasing, and satisfies~$h^{\|\cdot\|}(0)=0$ and $h^{\|\cdot\|}(+\infty)=+\infty$; these transformations extend those in~(\ref{univransf}) in a spherical fashion and form a group that generates the null~$\mathcal{H}_0^{\rm spher}$ of bivariate spherical symmetry about the origin. Since $\mathcal{H}_0^{\rm spher}\subset \mathcal{H}_0^{\rm centr}$, this implies that~$R\n_{D}$ (at least for the three depth functions considered above) cannot be invariant under a group of transformations that generates~$\mathcal{H}_0^{\rm centr}$. 

The statistics~$R\n_{D}$, however, are permutation-invariant and affine-invariant. Affine-invariance, which is a classical requirement in multivariate statistics,  removes any dependence on the choice of the underlying coordinate system and ensures that the performances of the corresponding tests will not be affected by the variance-covariance structure---under infinite second-order moments, the ``scatter" structure---of the underlying distribution.

\begin{Prop}\label{runsaffinv}
(i) $R^{(n)}_{D}$ is invariant under permutations of the observations. 
(ii) If~$D$ satisfies~P1 from Section~\ref{depthsec}, then $R^{(n)}_{D}$ is affine-invariant, in the sense that~$R^{(n)}_{D}(\Ab\Xb_1,\ldots,\Ab\Xb_n)
\linebreak
=R^{(n)}_{D}(\Xb_1,\ldots,\Xb_n)$ for any  invertible $2\times 2$ matrix~$\Ab$.
\end{Prop}


We omit a proof here, as Part (i) is obvious and Part~(ii) follows from both the affine-invariance of the anti-ranks $A_i$ (thanks to P1) and the affine-invariance of the indicator function of the event that the origin belongs to a data-based simplex (which can be established as in \cite[Page 407]{Liu1990}). This entails affine-invariance of~$R\n_{D^H}$,~$R\n_{D^S}$ and~$R\n_{D^{SV}_{\rm mod}}$. Note that, in order to ensure affine-invariance of the anti-ranks, it suffices that the center-outward ordering is affine-invariant, while the exact value of the depth needs not be affine-invariant. This shows that $R^{(n)}_{D^{SV}}$ is affine-invariant, too.

\section{Asymptotic null distribution}\label{runsdist}

In this section, we derive the asymptotic null distribution of~$R^{(n)}_{D}$, which is of course needed to apply the corresponding test~$\phi_D^{(n)}$ (at a fixed asymptotic level~$\alpha$). 
Obtaining this asymptotic distribution of~$R\n_D$, however, is much more difficult than deriving the asymptotic null distribution of the McWilliams' test statistic, as, unlike the summands in the latter, the summands in~$R\n_D$ are not mutually independent. Note that they further do not form a stationary sequence. We therefore need a nonstandard CLT, that also applies to triangular arrays of random variables (since the whole collection of anti-ranks may be affected by the introduction of an extra observation~$\Xb_{n+1}$). We will make use of the following recent result.

%

\begin{theor}[\cite{Neu2013}] \label{theorhoeff}
Let $(Z_{n,i})_{i=1,\ldots,n}$, $n\in\N_0$, be a triangular array of random variables with mean zero. Assume that 
\begin{enumerate}
\item[(i)] $\sup_n\sum_{i=1}^n {\rm E}[Z_{n,i}^2]<\infty$;
\item[(ii)] for all~$\varepsilon>0$, $\sum_{i=1}^n {\rm E}[Z_{n,i}^2\mathbb{I}_{[|Z_{n,i}|>\varepsilon]}]=o(1)$ as~$n\to\infty$;
\item[(iii)] there exists a summable sequence $(a_{h})$ such that, for all~$m\in\N_0$ and all indices~$1\leq i_1<i_2<\ldots<i_m+h=:j_1\leq j_2\leq n$, 
$$
\big| {\rm Cov}[g(Z_{n,i_1},\ldots,Z_{n,i_m}),Z_{n,j_1}] \big|
\leq 
a_h ({\rm E}[g^2(Z_{n,i_1},\ldots,Z_{n,i_m})])^{1/2} \max(({\rm E}[Z_{n,j_1}^2])^{1/2},n^{-1/2})
$$
for all measurable and square integrable functions~$g:\R^m\to \R$, and
$$
\big| {\rm Cov}[g(Z_{n,i_1},\ldots,Z_{n,i_m}),Z_{n,j_1}Z_{n,j_2}] \big|
\leq 
a_h \|g\|_\infty \big( {\rm E}[Z_{n,j_1}^2] + {\rm E}[Z_{n,j_2}^2] + n^{-1}\big) 
$$
for all measurable and bounded functions~$g:\R^m\to \R$ with~$\|g\|_\infty:=\sup_{\xb\in\R^m}|g(\xb)|$.
\end{enumerate}
Then, provided that $\sigma^2:=\lim_{n\to\infty} {\rm Var}[\sum_{i=1}^n Z_{n,i}]<\infty$,  $\sum_{i=1}^n Z_{n,i}$ is asymptotically normal with mean zero and variance $\sigma^2$.
\end{theor}

In order to apply this result, we need the subsequent three lemmas (which are proved in the Appendix) and the two following assumptions:
\begin{itemize}
\item[(A1)] \emph{consistency}: $\sup_{\xb\in \R^2}|D(\xb,{\rm P}^{(n)})-D(\xb,{\rm P})|=o(1)$ almost surely as $n\to\infty$, where~${\rm P}\n$ denotes the empirical distribution associated with $n$ random vectors that are \mbox{i.i.d.}~${\rm P}$.
\item[(A2)] \emph{strict monotonicity}: the mapping 
$\alpha\mapsto g_{\rm P}(\alpha)= {\rm P}[\{\xb\in\R^2 : D(\xb,{\rm P})\geq \alpha\}]$ is strictly decreasing on~$(\alpha_{\rm min},\alpha_{\rm max})$, with $\alpha_{\rm min}=\inf\{\alpha >0 : g_{\rm P}(\alpha)<1\}$ and $\alpha_{\rm max}=\sup\{\alpha >0 : g_{\rm P}(\alpha)>0\}$.
\end{itemize}

Assumption~(A1) is satisfied for halfspace depth, simplicial depth, and projection depth (under mild assumptions on the univariate location and scale functionals used in this depth); see \cite{Zuo2003}, Remark~2.5. Under finite second-order moments, it also holds for Mahalanobis depth; see \cite{LiuSin1993}, Remark~2.2. As for Assumption~(A2), 
it is easy to show that it holds in particular when (i) ${\rm P}$ is absolutely continuous with respect to the Lebesgue measure over~$\R^2$, (ii) the support~$C$ of ${\rm P}$ is convex, and (iii) $\xb\mapsto D(\xb,{\rm P})$ is continuous\footnote{In the absolutely continuous case considered, continuity holds for most depths, including, e.g., halfspace depth, simplicial depth, and projection depth; see,  in \cite{PaiVan2013}, Assumption~(Q1), the comment below Theorem~3.1, and the proof of Lemma~A.1.}.

\begin{Lem}\label{lemruns1}
Let $\xb,\yb,\zb\in\R^2$ be in ``general position from the origin"---in the sense that all straight lines through the origin contain at most one element of~$\{\xb,\yb,\zb\}$. Then there are  exactly two vectors $(s_x,s_y,s_z)\in\{-1,1\}^{3}$ such that $\zerob\in S(s_x\xb,s_y\yb,s_z\zb)$, and those two vectors are opposite of each other. 
\end{Lem}


\begin{Lem}\label{lemrunsnew}
Let $\Xb_1,\ldots,\Xb_4$ be \mbox{i.i.d.} random vectors in $\R^2$ with common centrally symmetric distribution $\rm P$. Then, for any~$\tau\in(0,\sup_\xb D(\xb,P))$, the probability  
\begin{eqnarray*}
p_{\tau,{\rm P}}
&=&
{\rm P}
\Big[
\zerob\in S(\Xb_{1},\Xb_{2},\Xb_{3}),
\zerob\in S(\Xb_{2},\Xb_{3},\Xb_{4})
\,|\,
D(\Xb_i,{\rm P})=\tau, 
\
i=1,2,3,4
\Big]
\\
&=&
{\rm E}\Big[
\mathbb{I}_{\displaystyle [\zerob\in S(\Xb_{1},\Xb_{2},\Xb_{3})]}
\mathbb{I}_{\displaystyle [\zerob\in S(\Xb_{2},\Xb_{3},\Xb_{4})]}
\,|\,
D(\Xb_i,{\rm P})=\tau, 
\
i=1,2,3,4
\Big]
\end{eqnarray*}
is equal to $\frac{1}{12}$.
\end{Lem}
\vspace{1mm}

\begin{Lem}\label{lemruns3}
Let Assumptions~(A1)-(A2) hold, and consider the triangular array of random variables $(\mathbb{I}_{n,i})_{i=3,\ldots,n}$, $n\in \{3,4,\ldots\}$, where $\mathbb{I}_{n,i}:=\mathbb{I}_{\displaystyle [\zerob\in S(\Xb_{A_i},\Xb_{A_{i-1}},\Xb_{A_{i-2}})]}$. Then, under $\mathcal{H}_0^{\rm centr}$, 
(i)~${\rm E}[\mathbb{I}_{n,i}]=1/4$ for all $n,i$;
(ii)
for any~$\rho\in(0,1/2)$, 
$$
\sup_{i\in\mathcal{I}_{n}(\rho)}
\left|
{\rm E}[\mathbb{I}_{n,i}\mathbb{I}_{n,i-1}]
-
\frac{1}{12}
\right|
=o(1)
$$
as~$n\to\infty$, where~$\mathcal{I}_{n}(\rho):=\{\lfloor \rho n\rfloor,\lfloor \rho n\rfloor+1,\ldots,\lfloor (1-\rho)n\rfloor\}$;
(iii) for all~$n$, the sequence $(\mathbb{I}_{n,i})_{i=3,\ldots,n}$ is $1$-dependent.
\end{Lem}

%

With this in hand, we can then state the main result of this section; see the Appendix for a proof.

\begin{theor}\label{Rnasymp}
Let Assumptions~(A1)-(A2) hold, and assume that the statistical depth function~$D$ satisfies P2-P4. Then,
under $\mathcal{H}_0^{\rm centr}$, 
$
n^{-1/2}\big(4R^{(n)}_{D}-n-2\big)
$
is asymptotically normal with mean zero and variance $\sigma^2=11/3$.
\end{theor}

This theorem shows that the statistics~$R^{(n)}_{D}$ are asymptotically distribution-free. Of course, it also implies that the resulting tests~$\phi^{(n)}_{D}$ reject the null  of central symmetry~$\mathcal{H}_0^{\rm centr}$ at asymptotic level~$\alpha$ whenever
$$
\frac{4R^{(n)}_{D}-n-2}
{\sqrt{11n/3}}
<\Phi^{-1}(\alpha),
$$
where $\Phi$ stands for the cumulative distribution function of the standard normal distribution.

\section{Monte Carlo experiments}\label{runssimus}

The aim of this section is to conduct a Monte Carlo study that evaluates the finite-sample performances of the proposed tests. 
We start by describing briefly the competing procedures we will consider.

\subsection{Some competitors to our runs tests}\label{runssimustest}

We consider nine competitors, which may be grouped into the following four classes:
\begin{itemize}
\item[$\bullet$]
The first competitors are related to the runs tests for spherical symmetry proposed in \cite{Mar1999}. His extension of the \cite{McW1990} procedure consists in reordering (Step~(i)) the observations $\Xb_1,\ldots,\Xb_n$ as $\Xb_{A_1},\ldots,\Xb_{A_n}$ according to their Euclidean norms~$\|\Xb_i\|$, and then defining (Step~(ii)) his bivariate runs as consecutive inner products in the series~$\Ub_{A_i}$, $i=1,\ldots,n$, where $\Ub_{i}:=\Xb_{i}/\|\Xb_i\|$ is the so-called \emph{spatial sign} of~$\Xb_i$. More precisely, \cite{Mar1999}'s bivariate runs test is based on the statistic
$$
T^{(n)}_{\rm Marden}
=
\sqrt{\frac{2}{n}}\,
\sum_{i=2}^n\Ub_{A_i}'\Ub_{A_{i-1}},
$$
which is asymptotically standard normal under $\mathcal{H}_0^{\rm spher}$, the null hypothesis of spherical symmetry about the origin. Besides the one-sided test $\phi^{(n){\rm spher}}_{\rm Marden1}:=\mathbb{I}_{[T^{(n)}_{\rm Marden}>\Phi^{-1}(1-\alpha)]}$, which is a natural extension of the univariate \cite{McW1990} test, we also consider the two-sided test $\phi^{(n){\rm spher}}_{\rm Marden2}:=\mathbb{I}_{[(T^{(n)}_{\rm Marden})^2>\chi^2_{1,1-\alpha}]}$ (where~$\chi^2_{\ell,1-\alpha}$ denotes the $\alpha$-upper quantile of the $\chi^2_\ell$ distribution), as this is actually the test described in \cite{Mar1999}. 

The use of Euclidean distances leaves no doubt about the spherical nature of these tests. However, it is possible to extend them into tests of elliptical symmetry about the origin. Such tests are obtained by applying Marden's tests on standardized observations~$\hat{\pmb\Sigma}^{-1/2}\Xb_i$, $i=1,\ldots,n$, where~$\hat{\pmb\Sigma}$ is some affine-equivariant shape estimator---in the sense that  for any invertible $2\times 2$ matrix~$\Ab$, $\hat{\pmb\Sigma}(\Ab\Xb_1,\ldots,\Ab\Xb_n)=c\Ab\hat{\pmb\Sigma}\Ab^\prime$ for some constant~$c$ that may depend on the sample. Below we use the \cite{Tyl1987} shape estimator (with fixed location~$\mathbf{0}\in\R^2$), which is defined as the solution of $\frac{1}{n} \sum_{i=1}^n\Xb_i \Xb_i'/(\Xb_i'  {\pmb{\Sigma}}^{-1} \Xb_i)=\frac{1}{2}{\pmb{\Sigma}}$ under the constraint ${\rm Trace}[{\pmb{\Sigma}}]=2$. This leads us to add the ellipticity tests~$\phi^{(n){\rm ellipt}}_{\rm Marden1}$ and $\phi^{(n){\rm ellipt}}_{\rm Marden2}$ to our simulation study. Since the emphasis of our work is not on these ellipticity tests, we do not prove that their asymptotic distributions under $\mathcal{H}_0^{\rm ellipt}$ (the hypothesis of elliptical symmetry about the origin) do coincide  with those of $\phi^{(n){\rm spher}}_{\rm Marden1}$ and $\phi^{(n){\rm spher}}_{\rm Marden2}$ under $\mathcal{H}_0^{\rm spher}$ (yet the simulations below suggest that this is indeed the case).

\item[$\bullet$] \cite{Bar1991} proposes a class of sphericity tests based on statistics of the form
$$B^{(n)}=\frac{1}{n}\sum_{i,j=1}^nh(\Ub_i'\Ub_j)\min\left(1-\frac{R_i-1}{n},1-\frac{R_j-1}{n}\right),$$
where~$h$ is defined over~$[-1,1]$ and satisfies some regularity conditions (see Baringhaus~1991 for more details)  and where  $R_i$, $i=1,\ldots,n$, is the rank of~$\|\Xb_i\|$ among~$\|\Xb_1\|,\ldots,\|\Xb_n\|$. We restrict below to $h(t)=\big(t-\frac{1}{4}\big)/\big(\frac{17}{8}-t\big)$,  $t\in[-1,1]$, for mainly two reasons: (i) the asymptotic null distribution of $B^{(n)}$ then coincides (up to a multiplicative constant) with that of (the squared of) the natural Kolmogorov-Smirnov statistic for the problem under study, hence is fairly standard (whereas other choices of~$h$ would necessitate simulations to approximate the limiting null distribution); (ii) the resulting test---$\phi^{(n){\rm spher}}_{\rm Bar}$, say---then is a universally consistent sphericity test. 

Again, we also consider the extension of this sphericity test into an ellipticity test, $\phi^{(n){\rm ellipt}}_{\rm Bar}$ say (still obtained by applying the test $\phi^{(n){\rm spher}}_{\rm Bar}$ to observations~$\hat{\pmb\Sigma}^{-1/2}\Xb_i$, $i=1,\ldots,n$, standardized through Tyler's estimator of shape; here as well, our simulations tend to confirm that $\phi^{(n){\rm ellipt}}_{\rm Bar}$ is a valid ellipticity test).

\item[$\bullet$] We further consider the pseudo-Gaussian ellipticity test~$\phi^{(n){\rm ellipt}}_{\rm Cassart}$ described in \cite{Cas2007}, Chapter~3. This test achieves Le Cam optimality against the Fechner-type multinormal alternatives defined therein. It rejects $\mathcal{H}_0^{\rm ellipt}$  at asymptotic level~$\alpha$ whenever 
$$\frac{8}{3nm^{(n)}_4}\sum_{i,j=1}^n \hat d_i^2 \hat d_j^2\SSbb'_{\hat{\Ub}_i}\SSbb_{\hat{\Ub}_j}
>\chi^2_{2,1-\alpha},$$ 
where $\hat d_i:=(\Xb_i'\hat{\Sigmab}^{-1}\Xb_i)^{1/2}$, $\hat{\Ub}_i:=\hat{\Sigmab}^{-1/2}\Xb_i/\hat d_i$, $\SSbb_{\hat{\Ub}_i}:=(((\hat{\Ub}_i)_1)^2{\rm Sign}((\hat{\Ub}_i)_1),((\hat{\Ub}_i)_2)^2
\linebreak
{\rm Sign}((\hat{\Ub}_i)_2))'$, and $m^{(n)}_4:=\frac{1}{n}\sum_{i=1}^n \hat d_i^4$. We still use Tyler's estimator of shape for~$\hat{\pmb\Sigma}$. This parametric test requires finite fourth-order moments.

\item[$\bullet$] Finally, we also consider the \cite{Blo1989} projection pursuit tests for central symmetry. These tests first apply some univariate symmetry test~$\phi_{\rm univ}\n$ on the projected data set~$\ub_1'\Xb_i$, $i=1,\ldots,n$, with
$$
\ub_1
:=
\arg\min_{\ub\in\mathcal{S}^1}
\bigg[
\max_{1\leq i\leq n}\frac{(\ub'\Xb)_{(i)}+(\ub'\Xb)_{(n-i+1)}}{2}
-
\min_{1\leq i\leq n}\frac{(\ub'\Xb)_{(i)}+(\ub'\Xb)_{(n-i+1)}}{2}
\bigg]
,
$$
where $(\ub'\Xb)_{(j)}$ stands for the $j$th order statistic of the projected sample $\ub'\Xb_i$, $i=1,\ldots,n,$ and $\mathcal{S}^1$ is the unit circle in $\R^2$. If this univariate symmetry test rejects the null, then the projection pursuit test does reject~$\mathcal{H}_0^{\rm centr}$.  In case of no rejection, the same univariate test is run for the direction orthogonal to~${\ub}_1$. The projection pursuit test then does not reject~$\mathcal{H}_0^{\rm centr}$ iff this second run of~$\phi_{\rm univ}\n$ does not lead to rejection. If the overall test should have level~$\alpha$, then the two runs of~$\phi_{\rm univ}\n$ should  be conducted at level~$\alpha/2$ according to Bonferroni. For~$\phi_{\rm univ}\n$, we will use the classical test of skewness (based on empirical third-order moments) and the \cite{McW1990} runs test. The resulting projection pursuit tests will be denoted as $\phi^{(n)}_{\rm PPG}$ and $\phi^{(n)}_{\rm PPR}$, respectively.

\end{itemize}

\subsection{Finite-sample performances of our runs tests}\label{runssimusas}

In order to compare the finite-sample performances of the proposed tests with those of their nine competitors described above, we have considered several settings.

\begin{figure}
\begin{center}
\includegraphics[width=.95\textwidth]{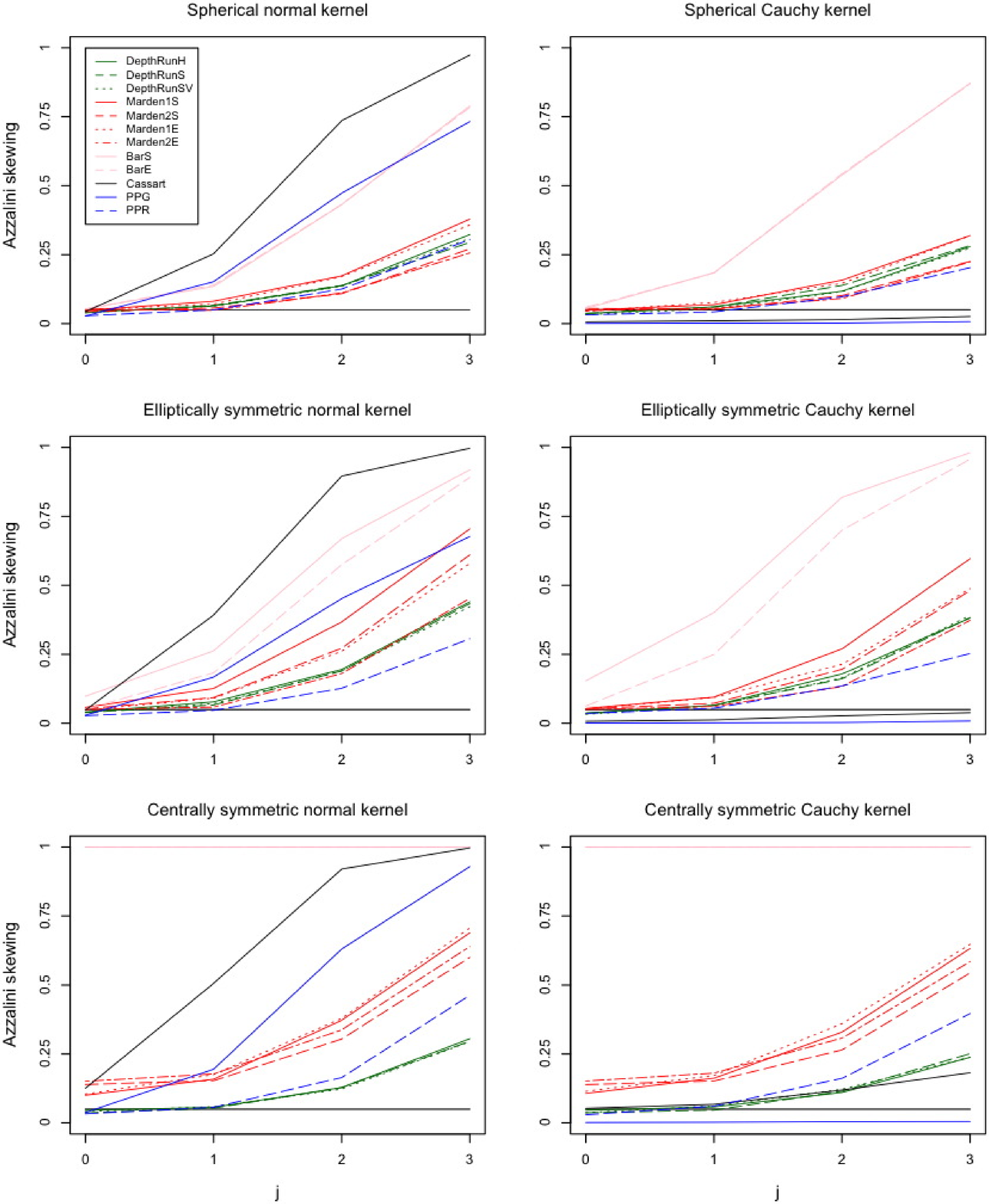}
\end{center}
\vspace{-6mm}
\caption{Curves of rejection frequencies (out of 3,000 replications), under skewed (by means of Azzalini-type skewing mechanisms) spherically, elliptically and centrally symmetric bivariate normal and Cauchy distributions, of the proposed depth-based runs tests, the \cite{Mar1999}-type tests, the \cite{Bar1991}-type tests, the pseudo-Gaussian test from \cite{Cas2007}, and the projection pursuit tests from \cite{Blo1989}, for samples of size~$n=100$; see Section~\ref{runssimus}  for details.}
      \label{asympsimus1}
\end{figure}

\begin{figure}
\begin{center}
\includegraphics[width=.95\textwidth]{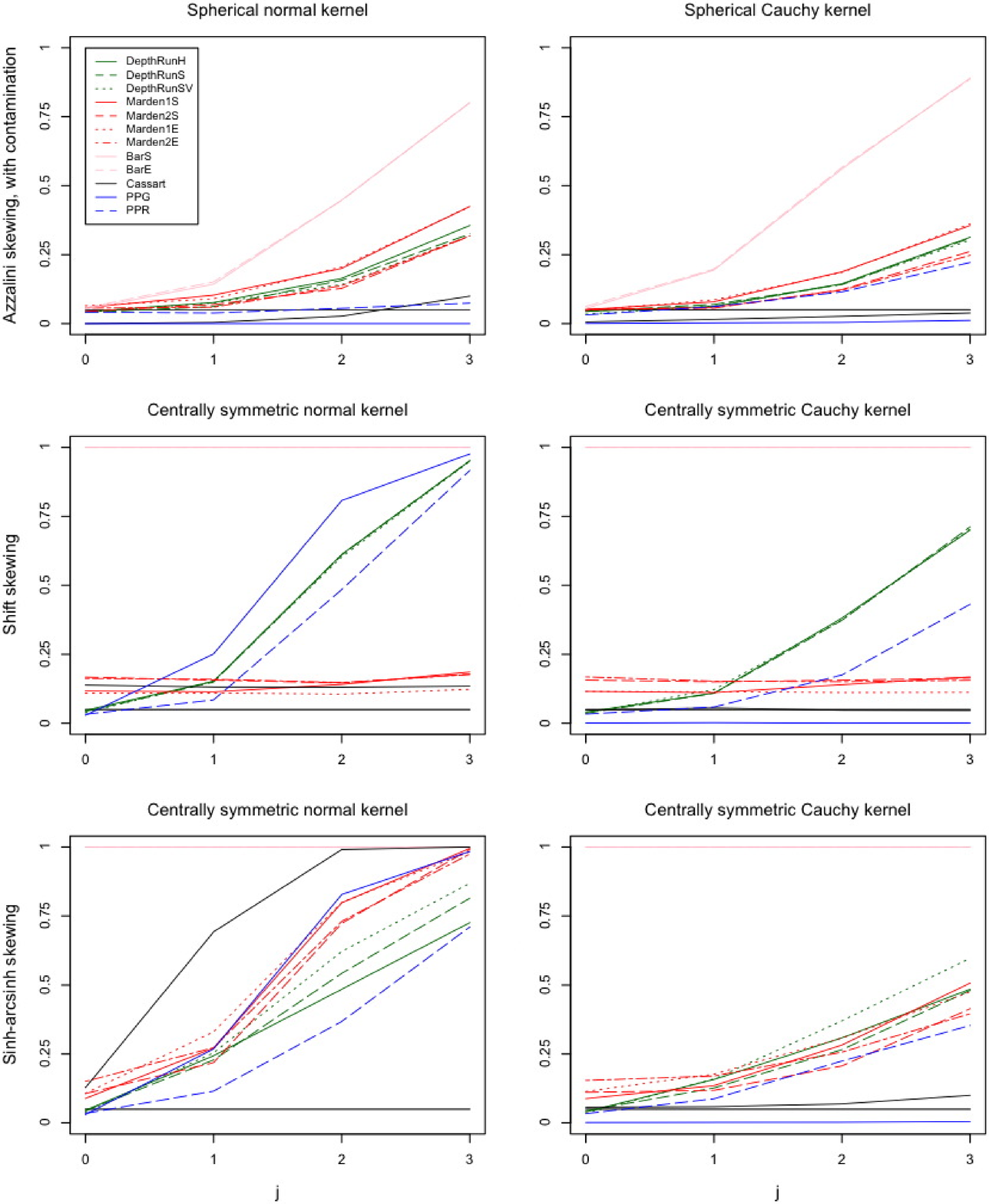}
\end{center}
\vspace{-6mm}
       \caption{Curves of rejection frequencies (out of 3,000 replications), under contaminated skewed  (by means of Azzalini-type skewing mechanisms) spherically symmetric distributions, shifted centrally symmetric distributions, and sinh-arcsinh-transformed centrally symmetric distributions (in each case with bivariate normal and Cauchy kernels), of the same tests as in Figure~\ref{asympsimus1} for samples of size~$n=100$; see Section~\ref{runssimus}  for details.}
       \label{asympsimus2}
\end{figure}

\begin{figure}
\begin{center}
\includegraphics[width=.46\textwidth]{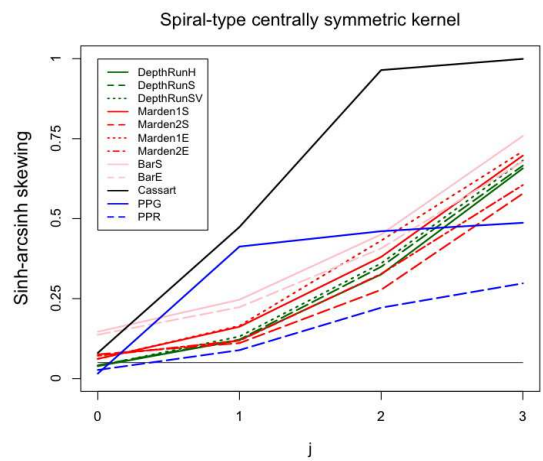}
\end{center}
\vspace{-6mm}
\caption{Curves of rejection frequencies (out of 3,000 replications), under a sinh-arcsinh-transformed centrally symmetric spiral-type distribution, of the same tests as in Figures~\ref{asympsimus1}-\ref{asympsimus2} for samples of size~$n=100$; see Section~\ref{runssimus} for details.}
\label{asympsimus3}
\end{figure}

 In each setting, $3,000$ independent random samples of size $n=100$ were generated from a centrally symmetric kernel (associated with $j=0$ below) and three increasingly skewed versions of this original symmetric distribution (associated with $j=1,2,3$ below) obtained from a particular skewing mechanism. Each sample  was subjected to the runs tests~$\phi^{(n)}_{D^{H}}$, $\phi^{(n)}_{D^S}$, and~$\phi^{(n)}_{D^{SV}}$ (based, respectively, on the halfspace, the simplicial and the simplicial volume depth) and to their nine competitors, all at nominal level~$5\%$. 
The resulting rejection frequencies are plotted against~$j$ in Figures~\ref{asympsimus1}-\ref{asympsimus3},  while Table~\ref{Tsimu1} contains the numerical values for two situations where the power curves present a strong overlapping.  Of course, the various settings differ by the symmetric kernels and/or the skewing mechanisms involved.

The six settings in Figure~\ref{asympsimus1} mainly differ by 
the symmetric kernels used. In the first (resp., second) row, these kernels are spherical (resp., elliptical) bivariate normal and Cauchy distributions, with shape parameter
$$
\Sigmab=
\Bigg(
\begin{array}{cc}
1 & 0 \\
0 & 1 
\end{array}
\Bigg)
\qquad
\Bigg(
 \textrm{resp., }
 \Sigmab=
\Bigg(
\begin{array}{cc}
2 & 1 \\
1 & 3 
\end{array}
\Bigg)
\Bigg)
.
$$
In the third row, we used centrally symmetric  kernels associated with the distributions obtained by conditioning bivariate spherical normal and Cauchy random vectors on the event that the random vectors belong to the (two-sided) cone~$\mathcal{C}_1:=\{ (x_1,x_2)' \in \R^2 : |\arctan (x_2/x_1)| \leq 1/2 \}$. 
For each of these six settings, the corresponding symmetrically distributed random vectors~$\Zb_1,\ldots,\Zb_n$ were skewed into $\Xb_1,\ldots,\Xb_n$ through
$$
\left\{
\begin{array}{lc}
\Xb_i=\Zb_i&\mbox{if}\,\,U\leq \Phi(j\deltab'\Zb_i)\\
\Xb_i=-\Zb_i&\mbox{if}\,\,U> \Phi(j\deltab'\Zb_i),
\end{array}\right. 
\qquad j=0,1,2,3,
$$
for normal kernels, and through
$$
\left\{
\begin{array}{lc}
\Xb_i=\Zb_i&\mbox{if}\,\,U\leq T_3(j\deltab'\Zb_i(3/(1+\Zb_i'\Zb_i))^{1/2})\\
\Xb_i=-\Zb_i&\mbox{if}\,\,U>T_3(j\deltab'\Zb_i(3/(1+\Zb_i'\Zb_i))^{1/2}),
\end{array}\right. 
\qquad j=0,1,2,3,
$$
for Cauchy ones, where $\deltab=(0.15,0.15)'$,  $U$ is uniformly distributed over $(0,1)$, and $T_3$ stands for the cdf of the univariate $t$ distribution with 3 degrees of freedom. These skewing mechanisms go back to \cite{AzzDal1996} for the normal case and to \cite{AzzCap2003} for the Cauchy case. 

Figure~\ref{asympsimus1} reveals that, in all settings, the behavior of our tests does not depend much on the depth function used. As expected, while the sphericity and ellipticity tests collapse under central symmetry, our tests and the projection pursuit tests still meet the $5\%$ nominal level constraint under such conditions.  As for the non-null behavior, our tests always detect asymmetry, irrespective of the shape or tail weight of the underlying distribution. Moreover, they perform well under heavy tails. For instance, under the skewed centrally symmetric Cauchy distribution (lower right picture in Figure~\ref{asympsimus1}), only $\phi^{(n)}_{\rm PPR}$ beats our tests (note that $\phi^{(n)}_{\rm PPG}$ has no power at all there, a feature common to all settings based on a  Cauchy distribution). 
Simulations based on $t_3$ instead of Cauchy distributions led to very similar results (except that $\phi^{(n){\rm ellipt}}_{\rm Cassart}$ exhibits some power  there), which is the reason why we do not include the corresponding plots here.


Parallel to Figure~\ref{asympsimus1}, the first (resp., second) column of Figure~\ref{asympsimus2} reports rejection frequencies under skewed normal (resp., Cauchy) distributions. In the first row, in which we intended to investigate the robustness properties of the various tests, we considered the same distributions as in the first row of Figure~\ref{asympsimus1}, but replaced the last two observations with the outlying values $(10,10)^\prime$ and $(11,1)^\prime$. This has a dramatic impact on $\phi^{(n){\rm ellipt}}_{\rm Cassart}$ and on the projection pursuit tests (mainly on~$\phi^{(n)}_{\rm PPG}$), while the other tests, including ours, are not much affected by this contamination. Since further simulations have actually revealed that other contaminations yield very similar results, our  tests, parallel to their classical univariate antecedent, enjoy good robustness properties. Let us now turn our attention to the second row of Figure~\ref{asympsimus2}. The symmetric kernels there are centrally symmetric normal and Cauchy densities, but with the thinner (two-sided) cone $\mathcal{C}_2:=\{ (x_1,x_2)' \in \R^2 : |\arctan (x_2/x_1)| \leq 1/5 \}$. Asymmetry is now introduced by transforming the corresponding symmetric random vectors~$\Zb_i$ into $\Xb_i=\Zb_i+j(0,0.04)'$, $j=0,1,2,3$, that is by simply shifting the~$\Zb_i$'s in the direction orthogonal to the axis of the cone. While it remains true that only tests designed for central symmetry meet the $5\%$ level constraint and that  $\phi^{(n)}_{\rm PPG}$ exhibits no power under skew-Cauchy distributions, it is  interesting to note that our three tests here clearly outperform the projection pursuit test $\phi^{(n)}_{\rm PPR}$. 
Finally, the third row of Figure~\ref{asympsimus2} uses again centrally symmetric distributions, but, instead of a single two-sided cone, we consider the union of two such cones, namely $\mathcal{C}_2\cup\mathcal{C}_3$, where $\mathcal{C}_3:=\{ (x_1,x_2)' \in \R^2 : |\arctan (x_2/x_1)-\pi/4| \leq 1/10 \}$. The skewing method employed is based on the \emph{sinh-arcsinh transform} from \cite{JonPew2009}, which turns the corresponding symmetrically distributed $\Zb_i$'s into
\begin{equation}\label{SA}
\Xb_i=
\Bigg(
\begin{array}{c}
\Xb_{i1}\\
\Xb_{i2}
\end{array}
\Bigg)
=
\Bigg(
\begin{array}{c}
\sinh (\sinh^{-1}(\Zb_{i1})+j\delta_1)\\
\sinh (\sinh^{-1}(\Zb_{i2})+j\delta_2)
\end{array}
\Bigg),\quad j=0,1,2,3, 
\end{equation}
with $(\delta_1,\delta_2)'=(0.12,0.1)'$. The same comments as above concerning the validity of the tests under the null and the power of some competitors under a Cauchy kernel still apply here. Quite interestingly, note that skewing this centrally symmetric normal distribution by means of the sinh-arcsinh transform is actually the only example where  our three tests do not perform in the exact same way. Moreover, the plots speak a clear language: our tests perform well in this setting too, especially under the skew-Cauchy version. And in the skew-normal setting, they are dominated by $\phi^{(n)}_{\rm PPG}$  only. Overall, Figure~\ref{asympsimus2} thus shows that our tests perform uniformly well, irrespective of the skewing methods used.


Finally, we considered a totally different setting, in which the symmetric kernel is the pdf of the bivariate ``spiraled" random vector
$$
\Bigg(
\begin{array}{c}
Z_1\\[-.5mm]
Z_2
\end{array}
\Bigg)
=
SU(1+10\,\theta)
\,
\Bigg(
\begin{array}{c}
\cos\theta\\[-.5mm]
\sin\theta
\end{array}
\Bigg),
$$
where $S=\pm1$ with respective probability $1/2$, $U\sim {\rm Unif}(0,1)$ and $\theta\sim{\rm Unif}(0,\pi)$, all three random variables being mutually independent. 
Asymmetry was introduced via the sinh-arcsinh transform defined in~(\ref{SA}), now with $(\delta_1,\delta_2)'=(0.2,0.15)'$. Figure~\ref{asympsimus3} reports the resulting rejection frequencies, and shows that, again,  only our tests and the projection pursuit tests are able to meet the nominal level constraint. Quite interestingly, our tests uniformly outperform the runs-based projection pursuit test~$\phi^{(n)}_{\rm PPR}$. Note that they beat the Gaussian projection pursuit test~$\phi^{(n)}_{\rm PPG}$ for severely asymmetric distributions only; recall, however, that~$\phi^{(n)}_{\rm PPG}$  is poorly robust to heavy tails. 

As a summary, this Monte Carlo study shows that  our depth-based runs tests, unlike most of their competitors, always meet the nominal level constraint, and that they always detect asymmetry, whatever the symmetric kernel or skewing mechanism used. Quite nicely, they moreover exhibit good robustness properties and often outperform most of their competitors, which is particularly remarkable for tests that extend to the multivariate setup a univariate universally consistent procedure.
Finally, we report that simulations for sample sizes $n=50$ and $n=200$ led to very similar results, which explains that we restricted to~$n=100$ above.

\section{Final comments}\label{runsFC}

In this final section, we shall briefly discuss some open problems and possible extensions related to the material presented in this paper.

In the univariate case, \cite{ModGas1996} propose a weighted version of the \cite{McW1990} test statistic. The same weighting scheme straightforwardly applies in the bivariate setup, yielding weighted (depth-based) runs statistics of the form
$$
R_{D;\omega}^{(n)}=1+\sum_{i=2}^n\omega_i\mathbb{I}_{[\zerob\in S(\Xb_{A_i},\Xb_{A_{i-1}},\Xb_{A_{i-2}})]},
$$
where the $\omega_i$'s are positive real weights. The choice of the weights being totally free, one can give more importance to the observations near the center of symmetry by choosing a monotone decreasing sequence of weights, whereas, on the contrary, a monotone increasing sequence of $\omega_i$'s allows to base the outcome of the tests more on the observations in the ``tails". In contrast with this, $R^{(n)}_D$ treats equally observations from various depth levels. Therefore, it is clear that, in some situations, using $R^{(n)}_{D;\omega}$ instead of $R^{(n)}_D$ may increase the power.

The proposed bivariate runs tests can quite naturally be extended into tests for $k$-variate central symmetry. First, the statistical depth functions from Section~\ref{depthsec} were indeed described for an arbitrary dimension $k$, which defines the corresponding anti-ranks. Second, the bivariate simplicial runs introduced in~(\ref{ffg}) readily generalize into
$$
1+\sum_{i=k+1}^n
\,
\mathbb{I}_{\displaystyle[\zerob\in S(\xb_i,\xb_{i-1},\ldots,\xb_{i-k})]},
$$
where $S(\xb_1,\xb_2,\ldots,\xb_{k+1})$ stands for the $k$-dimensional simplex with vertices $\xb_1,\xb_2,\ldots,\xb_{k+1}\in\R^k$. Consequently, the resulting depth-based runs tests would reject the null of $k$-variate central symmetry for small values of
$$
R^{(n)}_{D;k}=1+\sum_{i=k+1}^n\mathbb{I}_{[\zerob\in S(\Xb_{A_i},\Xb_{A_{i-1}},\ldots,\Xb_{A_{i-k}})]}.
$$
Deriving the asymptotic null distribution of this test requires extending Lemmas~\ref{lemruns1}-\ref{lemruns3} to the $k$-dimensional setup. This can be achieved fairly easily (in Lemma~\ref{lemruns3}, the sequence of indicator functions is then $(k-1)$-dependent with marginal expectation~$1/2^k$), except for Lemma~\ref{lemrunsnew} and, consequently, Lemma~\ref{lemruns3}(ii). Computing---or even only showing distribution-freeness of---the \mbox{$k-1$} probabilities involved in the $k$-variate version of that result would typically need ordering the directions of the observations. For dimensions~$k\geq 3$, this means that vectors of~$k-1\geq 2$ angles should be ranked, which adds further spice to the problem but is also a very delicate issue, and hence left for future research work.

Finally, it is natural to wonder whether the proposed tests for bivariate central symmetry inherit the universal consistency property from their univariate antecedent; recall indeed that \cite{Hen1993} proved that the \cite{McW1990} test is universally consistent under absolute continuity. Henze's proof actually identifies McWilliams' test as a two-sample location runs test and then exploits universal consistency properties of such runs tests. In the univariate case, the two samples are naturally made of (i) the original positive observations and (ii) the reflections (about the origin) of negative observations. In the bivariate case, however, infinitely many halflines from the origin can bear observations and such a two-sample structure does not exist. Extending the proof from \cite{Hen1993} is therefore extremely challenging. Yet, our tests exhibit some power in all setups considered in the Monte Carlo study above, and this universal consistency therefore remains an interesting open problem.

\vspace{0.6cm}

\noindent  ACKNOWLEDGEMENTS \normalsize
\vspace{2mm}

Christophe Ley thanks the Fonds National de la Recherche Scientifique, Communaut\'{e} francaise de Belgique, for financial support via a Mandat d'Aspirant and a Mandat de Charg\'{e} de Recherche FNRS. The research of Davy Paindaveine has been supported by an ARC grant of the Communaut\'{e} Fran\c{c}aise de Belgique and by the IAP research network grant P7/06 of the Belgian government (Belgian Science Policy). 

\appendix

\section{Appendix: technical proofs}

\noindent{\sc Proof of Lemma~\ref{lemruns1}.} Fix $s_z=1$ and consider the system of equations
\begin{equation}\label{bary}
\lambda_x s_x\xb+\lambda_ys_y\yb+\lambda_z\zb=\zerob,
\end{equation}
to be solved in~$(\lambda_x,\lambda_y,\lambda_z)\in(\R_0)^{3}$. The general position assumption  implies that each of the $4$ couples of signs~$(s_x,s_y)\in\{-1,1\}^2$ generates a solution $(\lambda_x,\lambda_y,\lambda_z)$ of~(\ref{bary}). Clearly, only one of those $4$ couples produces~$\lambda$'s that all share the same sign. For that~$(s_x,s_y)$, 
\begin{equation}\label{combill}
\xi_x s_x\xb+\xi_ys_y\yb+\xi_z\zb=\zerob,
\textrm{ with }
\Bigg(
\begin{array}{c}
\xi_x\\[-3.323mm]
\xi_y\\[-3.323mm]
\xi_z
\end{array}
\Bigg)
=
\frac{1}{\lambda_x+\lambda_y+\lambda_z}
\,
\Bigg(
\begin{array}{c}
\lambda_x\\[-3.323mm]
\lambda_y\\[-3.323mm]
\lambda_z
\end{array}
\Bigg)
\in[0,1]^3,
\end{equation}
which entails that $\zerob\in S(s_x\xb,s_y\yb,\zb)$. 
For all other couples, it is impossible (irrespective of the normalization) to make all coefficients of the linear combination~(\ref{combill}) be positive, so that the corresponding  simplices cannot contain the origin. 
The same reasoning applies to the case~$s_z=-1$, and obviously leads to one single couple~$(s_x,s_y)$ that is opposite to the couple identified for~$s_z=1$.
\cqfd
\vspace{3mm}

\begin{figure}
  \begin{center}
        \vspace{0pt}\includegraphics[width=15.7cm]{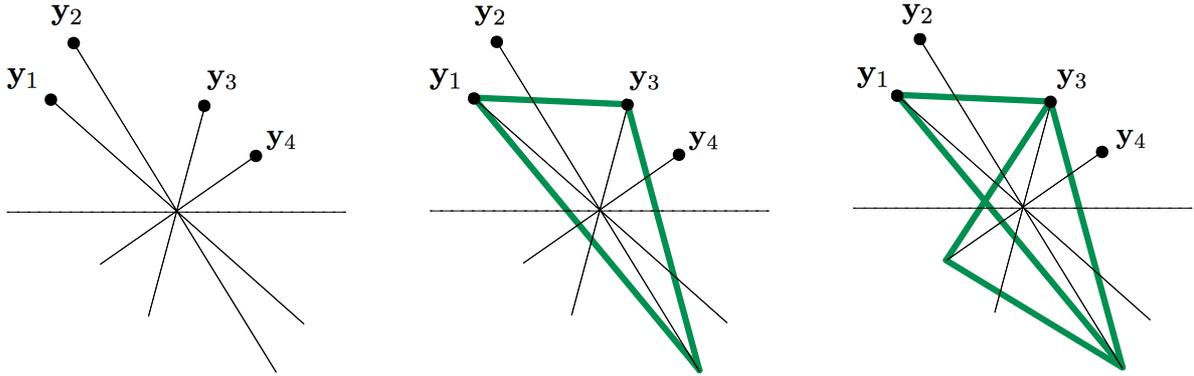}
    \caption{(left:) For the given $\yb$-configuration, exactly  two sign vectors $(s_1,s_2,s_3,s_4)\in\{-1,1\}^4$ are such that both simplices $S(s_1\yb_1,s_2\yb_2,s_3\yb_3)$ and $S(s_2\yb_2,s_3\yb_3,s_4\yb_4)$ contain the origin. To see this, first fix $s_1=1$.
    (center:) Identify then the unique (Lemma~\ref{lemruns1}) pair $(s_2,s_3)$ such that $S(\yb_1,s_2\yb_2,s_3\yb_3)$ contains the origin. (right:) Only one sign value~$s_4$ then provides a sign vector~$(1,s_2,s_3,s_4)$, namely $(1,-1,1,-1)$, for which both simplices contain the origin. Clearly, for $s_1=-1$, the same reasoning applies, resulting in the only sign vector $(s_1,s_2,s_3,s_4)=(-1,1,-1,1)$.}
    \label{11ok}
  \end{center}
\end{figure}

\begin{figure}
  \begin{center}
        \vspace{0pt}\includegraphics[width=15.7cm]{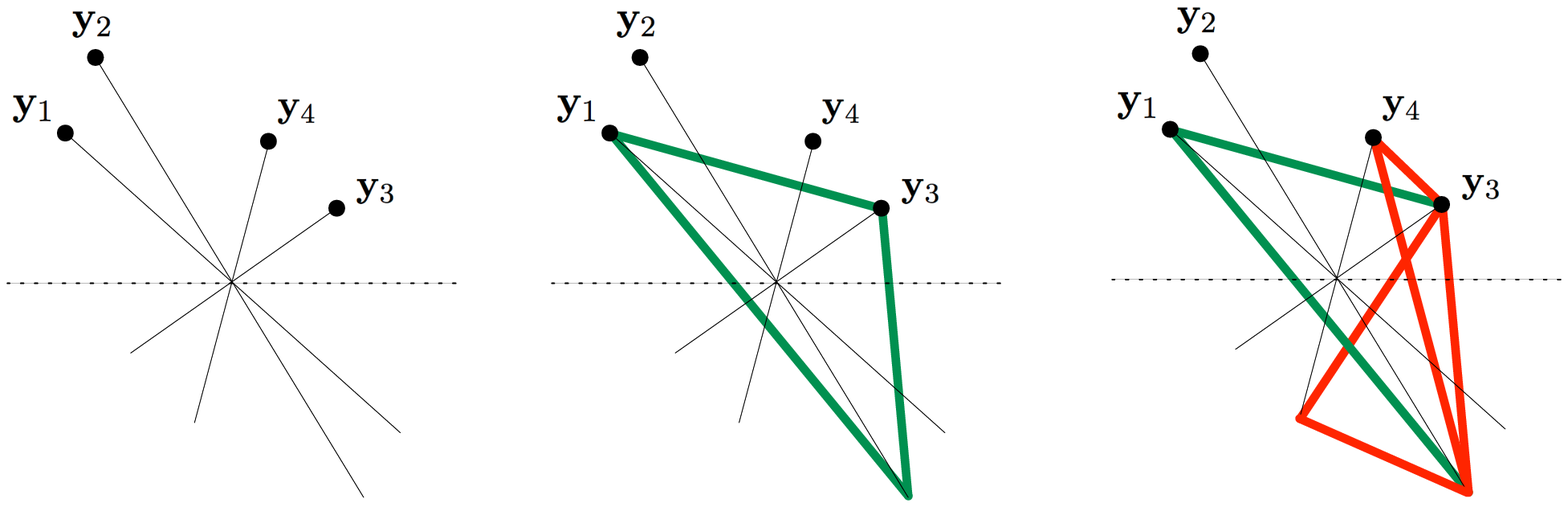}
    \caption{For the given $\yb$-configuration, 
    there is no sign vector $(s_1,s_2,s_3,s_4)\in\{-1,1\}^4$ for which both simplices $S(s_1\yb_1,s_2\yb_2,s_3\yb_3)$ and $S(s_2\yb_2,s_3\yb_3,s_4\yb_4)$ contain the origin (by proceeding as in~Figure~\ref{11ok}, it is seen that, in the right subfigure, no sign value~$s_4$ is such that  $S(s_2\yb_2,s_3\yb_3,s_4\yb_4)$ contains the origin).}\label{11notok}
  \end{center}
\end{figure}

\noindent{\sc Proof of Lemma~\ref{lemrunsnew}.} 
%
%
%
%
Let $\Yb:=(\Yb_1^{\prime},\Yb_2^{\prime},\Yb_3^{\prime},\Yb_4^{\prime})^\prime$, where~$\Yb_i:=S_i\Xb_i:={\rm Sign}((\Xb_i)_2)\Xb_i$, and fix  $\yb:=(\yb_1^{\prime},\yb_2^{\prime},\yb_3^{\prime},\yb_4^{\prime})^\prime\in (\partial D_{\tau}^+)^4$, with $\partial D_{\tau}^+:= \partial D_{\tau}\cap (\mathbb{R}\times\mathbb{R}^+)$, where $\partial D_{\tau}:=\{\xb\in\R^2\,:\, D(\xb,P)=\tau\}$. Since $\rm P$ is a centrally symmetric distribution, the signs $S_i$ are \mbox{i.i.d.} (they take values $\pm1$ with respective probability $1/2$) and are independent of~$\Yb$. 
It thus follows that 
\begin{eqnarray}
\lefteqn{
{\rm P}[\zerob\in S(\Xb_1,\Xb_2,\Xb_3), \zerob\in S(\Xb_2,\Xb_3,\Xb_4)\,|\,
D(\Xb_i,{\rm P})=\tau, \ i=1,2,3,4,
\, 
\Yb=\yb]}
\nonumber
\\
& &
\hspace{2mm}
={\rm P}[\zerob\in S(\Xb_1,\Xb_2,\Xb_3), \zerob\in S(\Xb_2,\Xb_3,\Xb_4)\,|\,
\Yb=\yb]
\nonumber
\\
& &
\hspace{2mm}
=\frac{1}{2^4}\sum_{(s_1,s_2,s_3,s_4)\in\{-1,1\}^4}
\mathbb{I}_{
\displaystyle{
[\zerob\in (S(s_1\yb_1,s_2\yb_2,s_3\yb_3)
}
{\displaystyle{\cap S(s_2\yb_2,s_3\yb_3,s_4\yb_4))]}
}}
%
=\frac{1}{2^4}\, \ell(\yb)
,
\nonumber
\label{seqequbiss}
\end{eqnarray}
where $\ell(\yb)$ denotes the number of sign vectors $(s_1,s_2,s_3,s_4)\in\{-1,1\}^4$ for which both $S(s_1\yb_1,s_2\yb_2,s_3\yb_3)$ and $S(s_2\yb_2,s_3\yb_3,s_4\yb_4)$ contain the origin. 
%
%
%
%
Lemma~\ref{lemruns1} implies that the only positive value of $l(\yb)$ is 2, but since we may actually have $l(\yb)=0$ (see Figures~\ref{11ok}-\ref{11notok} for an illustration), we may write
$$
{\rm P}[\zerob\in S(\Xb_1,\Xb_2,\Xb_3), \zerob\in S(\Xb_2,\Xb_3,\Xb_4)\,|\,
D(\Xb_i,{\rm P})=\tau, \ i=1,2,3,4,
\, 
\Yb=\yb]
=
\frac{1}{2^3}\,\mathbb{I}_{[\ell(\yb)=2]}
.
$$
Multiplying both sides of this equality with the density\footnote{With respect to the uniform distribution over~$(\partial D_{\tau}^+)^4$.} of $\Yb$ at $\yb$ conditional on $[D(\Xb_i,{\rm P})=\tau, \ i=1,2,3,4]$ (equivalently, conditional on~$[\Yb\in(\partial D_{\tau}^+)^4]$), and then integrating over $(\partial D_{\tau}^+)^4$ yields 
$$
\ptwoone_{\tau,{\rm P}}
=
\frac{1}{8}\, {\rm P}\big[\ell(\Yb)=2\,|\, \Yb\in(\partial D_{\tau}^+)^4\big]
.
$$
To evaluate this probability, we need to discriminate between the $\yb$'s in~$(\partial D_{\tau}^+)^4$ for which $\ell(\yb)=2$ and those for which $\ell(\yb)=0$. 

To this end, let $\theta_i:=\arccos((\Yb_i)_1/\|\Yb_i\|)$, $i=1,2,3,4$, be the angle between  the positive first semi-axis and the halfline~$\mathcal{L}_i:=\{\lambda \Yb_i: \lambda\geq 0\}$, and denote by $R_i$ the rank of $\theta_i$ among $\theta_1,\theta_2,\theta_3,\theta_4$. Clearly, $\ell(\Yb)$ is measurable with respect to~$(R_1,R_2,R_3,R_4)$. 
Now, among the $4!=24$ possible rankings $(R_1,R_2,R_3,R_4)$, it can easily be checked that exactly 16 are such that $\ell(\Yb)=2$ (these 16 rankings are made of the 12 rankings with $|R_2-R_3|=1$ and the 4 rankings with $\{R_2,R_3\}=\{1,4\}$). This, and the fact that, even conditional on~$[ \Yb\in(\partial D_{\tau}^+)^4]$, the 24 rankings are equally likely, eventually yields
$$
\ptwoone_{\tau,{\rm P}}
=\frac{1}{8}\, {\rm P}\big[\ell(\Yb)=2\,|\, \Yb\in(\partial D_{\tau}^+)^4\big]
=\frac{1}{8}\times\frac{16}{24}
=\frac{1}{12},
$$
as was to be proved. 
\cqfd
\vspace{3mm}

The proof of Lemma~\ref{lemruns3} requires the following technical result on depth, that is of independent interest.

\begin{Lem}
\label{Rainer}
Let~$D$ be a statistical  depth function and~$\rm P$ be a probability measure on~$\R^2$ that meet Assumptions~(A1)-(A2). Let $\Zb_1,\Zb_2,\ldots,\Zb_n$ be \mbox{i.i.d.} ${\rm P}$, and denote by~${\rm P}\n$ the resulting empirical distribution. Then $$\max_{i=2,\ldots,n} |D(\Zb_{(i)},{\rm P}\n)-D(\Zb_{(i-1)},{\rm P}\n)|\to 0$$ almost surely as~$\ny$, where $D(\Zb_{(i)},{\rm P}\n)$ denotes the $i$th order statistic of~$D(\Zb_1,{\rm P}\n),\ldots,D(\Zb_n,{\rm P}\n)$.   
\end{Lem}

\noindent{\sc Proof of Lemma~\ref{Rainer}.} 
Partition the interval $[\alpha_{\min},\alpha_{\max}]$ (see Assumption~(A2)) in~$M$ intervals
$K_{1,M},K_{2,M},\ldots,K_{M,M}$ of equal length. Since $g_{\rm P}$ is strictly decreasing on~$[\alpha_{\min},\alpha_{\max}]$, we have that 
$p_{m,M}:={\rm P}[D(\Zb_i,{\rm P})\in K_{m,M}]>0$ for all $m=1,\ldots,M$.
Denoting by $A_{m,M}^{(n)}$ the event that none of the $D(\Zb_i,{\rm P})$'s, $i=1,\ldots,n$, belong to~$K_{m,M}$, we have that, for all $M\in\N$ and $1\le m \le M$, 
$$
{\rm P}\big[A_{m,M}^{(n)}\big]
=
\big(
{\rm P}\big[ D(\Zb_1,{\rm P})\notin K_{m,M}\big]
\big)^n
=
(1-p_{m,M})^n
.
$$
so that, letting~$A_{m,M}:=\cap_{n=1}^\infty A_{m,M}^{(n)}$, we have that 
$$
{\rm P}[A_{m,M}]
=
\lim_{n\rightarrow\infty}{\rm P}\big[A_{m,M}^{(n)}\big]
=
0
.
$$ 
Since a countable union of null sets is again a null set, we conclude that
$$
{\rm P}\Bigg[\bigcup_{M=1}^\infty\bigcup_{m=1}^M A_{m,M}\Bigg]
=0
\,
;
$$
in other words, with probability 1, for any~$M\in\N$ and $m=1,\ldots,M$, there exists~$i\in\N_0$ such that  $D(\Zb_i,{\rm P})\in K_{m,M}$. Hence, for any~$M\in\N$, there exists~$N_M\in\N_0$ such that for all~$n\geq N_M$, each of the intervals~$K_{1,M},\ldots,K_{M,M}$ contains at least one of the depth values~$D(\Zb_1,{\rm P}),\ldots,D(\Zb_n,{\rm P})$. 

Now, let $\varepsilon>0$ and choose $M$ so large that the length --- $\delta$, say --- of the intervals $K_{1,M},\dots,K_{M,M}$ is smaller than $\varepsilon/4$. Then, from what we just have shown and by Assumption~(A1) it follows that, with probability~1, there exists $N$ such that for $n\ge N$, we have that 
(i)
$\sup_{\zb\in\R^2}|D(\zb\,|\,{\rm P}^{(n)})-D(\zb\,|\,{\rm P})|<\delta/2$, and
(ii)
 each of the intervals $K_{1,M},\dots,K_{M,M}$ contains at least 
one of the depth values $D(\Zb_1,{\rm P}),\dots,D(\Zb_n,{\rm P})$.

Fix then~$n\geq N$ and choose $i\in\{1,\ldots,n\}$ arbitrarily. From what precedes, we know that $D(\Zb_i,{\rm P})\in K_{m,M}$ for some 
$m\in\{1,\dots,M\}$. Then, two situations can occur:
\begin{itemize}
\item[1.] Assume that $m\geq3$. Pick then $j\in\{1,\ldots,n\}$ such that $D(\Zb_j,{\rm P})\in K_{m-2,M}$. We have that
$\delta\le D(\Zb_i,{\rm P}) - D(\Zb_j\,|\,{\rm P})\le 3\delta$. Since
$|D(\Zb_\ell\,|\,{\rm P}^{(n)})-D(\Zb_\ell\,|\,{\rm P})|<\delta/2$ for~$\ell=i,j$, we obtain
\begin{align*}
D(\Zb_i,{\rm P}^{(n)})-D(\Zb_j,{\rm P}^{(n)})
&>
\Big(D(\Zb_i,{\rm P})-\frac\delta2\Big)
-
\Big(D(\Zb_j,{\rm P})+\frac\delta2\Big)
\\
&=
D(\Zb_i,{\rm P})-D(\Zb_j,{\rm P}) - \delta 
\geq 0
\end{align*}
and
\begin{align*}
D(\Zb_i,{\rm P}^{(n)})-D(\Zb_j,{\rm P}^{(n)})
&<
\Big(D(\Zb_i,{\rm P})+\frac\delta2\Big)
-
\Big(D(\Zb_j,{\rm P})-\frac\delta2\Big)
\\
&=
D(\Zb_i,{\rm P})-D(\Zb_j,{\rm P}) + \delta
\le 4\delta
\le \varepsilon.
\end{align*}
\item[2.] Assume that $m\leq2$, that is, $D(\Zb_i,{\rm P})\in (K_{1,M}\cup K_{2,M})$ (which implies that $D(\Zb_i,{\rm P})\le \alpha_{\min}+2\delta$). 
Now, either there exists $j\in\{1,\ldots,n\}$ such that $D(\Zb_j,{\rm P}^{(n)})<D(\Zb_i,{\rm P}^{(n)})$, in which case 
\begin{align*}
D(\Zb_i,{\rm P}^{(n)})-D(\Zb_j,{\rm P}^{(n)})
&<
\Big(
D(\Zb_i,{\rm P})+\frac\delta2
\Big)
-
\Big(
D(\Zb_j,{\rm P})-\frac\delta2
\Big)
\\
&\le 
\Big(\alpha_{\min}+2\delta+\frac\delta2\Big) 
- 
\Big(\alpha_{\min}-\frac\delta2\Big)
=3\delta
<\varepsilon,
\end{align*}
or there is no such $j$, in which case $D(\Zb_i,{\rm P}^{(n)})=D(\Zb_{(1)},{\rm P}^{(n)})$.
\end{itemize}
Summing up, we have proved that, for every $i\in\{1,\dots,n\}$, either there exists  $j\in\{1,\dots,n\}\setminus\{i\}$ such that  
$0 < D(\Zb_i,{\rm P}^{(n)})-D(\Zb_j,{\rm P}^{(n)})<\varepsilon$  or $D(\Zb_i,{\rm P}^{(n)})= D(\Zb_{(1)},{\rm P}^{(n)})$. Clearly, this implies that   
$$
\max_{i=2,\dots,n}\left|D(\Zb_{(i)},{\rm P}^{(n)})-D(\Zb_{(i-1)},{\rm P}^{(n)})\right|<
\varepsilon
$$
for $n\geq N$, which establishes the result.
 \cqfd
\vspace{2mm}

We attract the reader's attention to the fact that, in  the previous proof, we have used ${\rm P}$ not only for the common density of the $\Zb_i$'s (as in the rest of the paper) but as well  for the probability measure of the underlying probability space on which the random quantities are defined (e.g., in ${\rm P}[A_{m,M}]$). This abuse of notation is voluntary in order to avoid unnecessarily complicated notations.

\vspace{2mm}

\noindent{\sc Proof of Lemma~\ref{lemruns3}.} 
(i) Define $\Yb:=(\Yb_1^{\prime},\ldots,\Yb_n^{\prime})^{\prime}$, with $\Yb_i:=S_i\Xb_i:={\rm Sign}((\Xb_i)_2)\Xb_i$, $i=1,\ldots,n$.
Note that the anti-ranks $A_i$, $i=1,\ldots,n$, are $\Yb$-measurable quantities, since they are computed on the basis of the  symmetrized sample. This, along with the fact that, under~$\mathcal{H}_0^{\rm centr}$, the $S_i$'s are \mbox{i.i.d.} (they take here as well values $\pm1$ with respective probability $1/2$) and are independent of~$\Yb$, yields
\begin{eqnarray}
{\rm E}[\mathbb{I}_{n,i}\, |\, \Yb]
& = & 
{\rm E}\big[\,\mathbb{I}_{\displaystyle [\mathbf{0}\in S(S_{A_i(\Yb)}\Yb_{A_i(\Yb)},S_{A_{i-1}(\Yb)}\Yb_{A_{i-1}(\Yb)},S_{A_{i-2}(\Yb)}\Yb_{A_{i-2}(\Yb)})]}\, |\,\Yb\big]
\nonumber
\\[1mm]
& = & 
 \frac{1}{2^{3}}
\sum_{(s_{i},s_{i-1},s_{i-2})\in\{-1,1\}^{3}}
\mathbb{I}_{\displaystyle [\mathbf{0}\in S(s_{i}\Yb_{A_i(\Yb)},s_{i-1}\Yb_{A_{i-1}(\Yb)},s_{i-2}\Yb_{A_{i-2}(\Yb)})]}.
\nonumber
\end{eqnarray}
In view of the absolute continuity of the~$\Yb_i$'s, Lemma~\ref{lemruns1} therefore shows that
\begin{equation} \label{fsjt}
{\rm E}[\mathbb{I}_{n,i} \,|\, \Yb]
=\frac{2}{2^{3}}
=\frac{1}{4}
\quad
\textrm{a.s.} 
\end{equation}
Taking expectations then yields the result.

(ii) Let $
\Delta_n=\max_{i=4,\ldots,n} ( \alpha_{n,i-3} - \alpha_{n,i})$, where~$\alpha_{n,i}=D(\Xb_{A_i},{\rm P}^{(n)}_{\rm sym})$, $i=1,\ldots,n$, by construction, forms a monotone decreasing sequence. For any~$i=4,\ldots,n$ and $k=0,1,2,3$, we then have that 
(i) 
$
\alpha_{n,i-k}
\leq \alpha_{n,i-3}< \alpha_{n,i-3}(1+1/n)
$
and
(ii) 
$
\alpha_{n,i-k}
\geq
\alpha_{n,i}
\geq
\alpha_{n,i-3} 
-
\Delta_n
$.
In other words, 
\begin{equation*}
\label{ouff}
\Xb_{A_i}\n,\Xb_{A_{i-1}}\n,\Xb_{A_{i-2}}\n,\Xb_{A_{i-3}}\n \in D\n_{\alpha_{n,i-3}-\Delta_n} \setminus D\n_{\alpha_{n,i-3}(1+1/n)}
\qquad
\forall i=4,\ldots,n,
\end{equation*}
where $D\n_\alpha$ denotes the collection of points~$\xb\in\R^2$ such that $D(\xb,{\rm P}^{(n)}_{\rm sym})\geq \alpha$.

Now, fix~$\rho\in(0,1/2)$, and restrict to indices~$i\in \mathcal{I}_{n}(\rho):=\{\lfloor \rho n\rfloor,\lfloor \rho n\rfloor+1,\ldots,\lfloor (1-\rho)n\rfloor\}$. With probability one, this ensures that, for $n$ large enough, $$
\Big[
\alpha_{n,\lfloor (1-\rho)n\rfloor-3}
,
\alpha_{n,\lfloor \rho n\rfloor-3}
\Big]
\subset
\Big[
\alpha(1-\rho/2)
,
\alpha(\rho/2)
\Big],
$$ 
where~$\alpha(\beta)$ is defined through~${\rm P}[D(\Xb,P)\geq \alpha(\beta)]=\beta$. Theorem~4.1 from \cite{ZuoSer2000B} implies that, for any~$\alpha\in[\alpha(1-\rho/2),\alpha(\rho/2)]$, $D\n_{\alpha-h_n}\setminus D\n_{\alpha}$ converges almost surely to~$\partial D_{\alpha}=\{\xb\in\R^2\,:\, D(\xb,P)=\alpha\}$ (recall that~${\rm P}^{(n)}_{\rm sym}$ converges weakly to~${\rm P}$ since~${\rm P}$ is symmetric about the origin). It is easy to check that the proof given in \cite{ZuoSer2000B} actually shows that this result also holds uniformly in~$\alpha\in[\alpha(1-\rho/2),\alpha(\rho/2)]$. This uniform convergence and Lemma~\ref{Rainer} (which implies that $\Delta_n$ converges to zero almost surely as~$\ny$), along with Lemma~\ref{lemrunsnew}, establishes the result.  
 
(iii) 
First note that~(\ref{fsjt}) yields
${\rm P}[\mathbb{I}_{n,i}=a\,|\,\Yb]=3^{1-a}/4(={\rm P}[\mathbb{I}_{n,i}=a])$ \mbox{a.s.}, $a\in\{0,1\}$, which implies that conditional (on $\Yb$) independence between $\mathbb{I}_{n,i}$ and $\mathbb{I}_{n,j}$ can be written as
\begin{equation}
{\rm P}[\mathbb{I}_{n,i}=a, \mathbb{I}_{n,j}=b\,|\,\Yb]
=
\frac{3^{1-a}}{4}\times\frac{3^{1-b}}{4},
\quad
a,b\in\{0,1\}.
\label{tosh2}
\end{equation}
Would this conditional independence hold true for $|i-j|\geq 2$, the lemma would follow since this would provide 
$$
{\rm P}[\mathbb{I}_{n,i}=a, \mathbb{I}_{n,j}=b]
=
{\rm E}[ {\rm P}[\mathbb{I}_{n,i}=a, \mathbb{I}_{n,j}=b\,|\,\Yb] ]
=
\frac{3^{1-a}}{4}\times\frac{3^{1-b}}{4}
={\rm P}[\mathbb{I}_{n,i}=a]{\rm P}[\mathbb{I}_{n,j}=b],
$$
 for all $a,b\in\{0,1\}$. 

We therefore conclude the proof by establishing the conditional independence above for $|i-j|\geq 2$. First, for $|i-j|\geq~3$, we can see that $\mathbb{I}_{n,i}=\mathbb{I}_{[\mathbf{0}\in S(S_{A_i(\Yb)}\Yb_{A_i(\Yb)},S_{A_{i-1}(\Yb)}\Yb_{A_{i-1}(\Yb)},S_{A_{i-2}(\Yb)}\Yb_{A_{i-2}(\Yb)})]}$ and $\mathbb{I}_{n,j}=\mathbb{I}_{[\mathbf{0}\in S(S_{A_j(\Yb)}\Yb_{A_j(\Yb)},S_{A_{j-1}(\Yb)}\Yb_{A_{j-1}(\Yb)}, S_{A_{j-2}(\Yb)}\Yb_{A_{j-2}(\Yb)})]}$ involve disjoint triples of signs, so that the result trivially follows from the mutual independence of those two collections of signs under the null. We may therefore focus on the case $|i-j|= 2$ (for which exactly one sign is common to both~$\mathbb{I}_{n,i}$ and $\mathbb{I}_{n,j}$). There, conditioning with respect to that common sign and then applying Lemma~\ref{lemruns1} to each corresponding simplex yields~(\ref{tosh2}) (after some immediate manipulations).
\cqfd
\vspace{3mm}

\noindent{\sc Proof of Theorem~\ref{Rnasymp}.} 
The strategy consists in applying Theorem~\ref{theorhoeff} to the statistic
$
(n-2)^{-1/2}\big(R^{(n)}_{D}-1-(n-2)/4\big)=\sum_{i=2}^{n}Z_{n,i},
$
based on the triangular array~$(Z_{n,i})_{i=3,\ldots,n}$, $n\in\{3,4,\ldots\}$, with~$Z_{n,i}:=(\mathbb{I}_{n,i}-\frac{1}{4})/\sqrt{n-2}$. 

Lemma~\ref{lemruns3}(i) directly shows that~${\rm E}[Z_{n,i}]=0$ for all~$n,i$, and that  
${\rm Cov}[Z_{n,i},Z_{n,i-h}]=(n-2)^{-1}{\rm Cov}[\mathbb{I}_{n,i},\mathbb{I}_{n,i-h}]=(n-2)^{-1}({\rm E}[\mathbb{I}_{n,i}\mathbb{I}_{n,i-h}]-\frac{1}{16})$ for $h=0,1$ (in particular, ${\rm E}[Z_{n,i}^2]={\rm Var}[Z_{n,i}]=3/[16(n-2)]$), while Lemma~\ref{lemruns3}(iii) yields ${\rm Cov}[Z_{n,i},Z_{n,i-h}]=0$ for $h\geq 2$. 
For~$n$ large (more precisely, $n\geq 4$), we therefore have
\begin{eqnarray}
\lefteqn{
\sigma^2_n
:=
{\rm Var}\bigg[\sum_{i=3}^n Z_{n,i}\bigg]
=
\frac{3}{16} 
+ 
\frac{2}{n-2}\, \sum_{i=4}^{n}\, ({\rm E}[\mathbb{I}_{n,i}\mathbb{I}_{n,i-1}]-1/16)
}
\nonumber
\\[2mm]
& &
\hspace{8mm}
=
\frac{3}{16} 
- 
\frac{2(n-3)}{16(n-2)} 
+ 
\frac{2}{n-2}\, \sum_{i=4}^{n}\, {\rm E}[\mathbb{I}_{n,i}\mathbb{I}_{n,i-1}]
\nonumber
\to
\frac{1}{16} 
+ 
2c
=:\sigma^2
\,
,
\nonumber
\end{eqnarray}
where we let
$$
c:=\lim_{n\to\infty} c_{n} :=\lim_{n\to\infty}
\Bigg[ \frac{1}{n-2}\,\sum_{i=4}^{n}\, {\rm E}[\mathbb{I}_{n,i}\mathbb{I}_{n,i-1}] \Bigg].
$$
To determine~$c$, we split~$c_{n}$ into
\begin{eqnarray}
c_{n}
&\!\!=\!\!&
c_{n,\rho}^{(1)}
+
c_{n,\rho}^{(2)}
+
c_{n,\rho}^{(3)}
\nonumber
\\[2mm]
&\!\!:=\!\!&
\frac{1}{n-2}\sum_{i=4}^{\lfloor \rho n\rfloor} \,{\rm E}[\mathbb{I}_{n,i}\mathbb{I}_{n,i-1}] 
+
\frac{1}{n-2}\sum_{i=\lfloor \rho n\rfloor+1}^{\lfloor (1-\rho)n\rfloor} {\rm E}[\mathbb{I}_{n,i}\mathbb{I}_{n,i-1}]
+
\frac{1}{n-2}\sum_{i=\lfloor (1-\rho)n\rfloor+1}^{n} {\rm E}[\mathbb{I}_{n,i}\mathbb{I}_{n,i-1}]
, 
\nonumber
\end{eqnarray}
where~$\rho\in(0,1/2)$ is fixed. It follows from Lemma~\ref{lemruns3}(ii) that
$
c_{n,\rho}^{(2)}
\to
(1-2\rho)/12
$
as $n\to\infty$, whereas the  inequality $|\mathbb{I}_{n,i}\mathbb{I}_{n,i-1}|\leq 1$ 
yields
$$
|c_{n,\rho}^{(1)}|
\leq 
\frac{\lfloor \rho n\rfloor-3}{n-2}
\to \rho
\quad \textrm{and}\quad
|c_{n,\rho}^{(3)}|
\leq 
\frac{n-\lfloor (1-\rho)n\rfloor}{n-2}
\to \rho
$$
as $n\to\infty$. 
Since this holds for any~$\rho\in(0,1/2)$, we conclude that $c=1/12$. This establishes the result since it is easy to check that the conditions~(i)-(iii) of Theorem~\ref{theorhoeff} hold ((i) is a trivial consequence of the identity~${\rm E}[Z_{n,i}^2]=3/[16(n-2)]$, (ii) follows from the boundedness of~$(n-2)^{1/2}|Z_{n,i}|$, whereas (iii) is a direct corollary of Lemma~\ref{lemruns3}(iii)).
%
\cqfd



%
%
%
%
%
%
%
%
%
%
%
%
%
%
%

\begin{table}
{
\begin{tabular}{@{}l|cccc}

 &  \multicolumn{4}{c} {Spherical normal kernel $\&$ Azzalini skewing with contamination}  \\[.5mm]

Test    &     $\deltab=(0,0)$    &    $\deltab=(0.15,0.15)$     &     $\deltab=(0.30,0.30)$  & $\deltab=(0.45,0.45)$      \\
\hline
DepthRunH & 0.0427   & 0.0767   & 0.1640   & 0.3557    	   \\
DepthRunS & 0.0403  &  0.0617   & 0.1570   & 0.3253      	  \\
DepthRunSV &     0.0413  &  0.0753  &  0.1413   & 0.3187     	   \\
BarS & 0.0527  &  0.1427   & 0.4477  &  0.8010            		    \\
Cassart&       0.0000  &  0.0047   & 0.0270  &  0.0997              	   \\

PPG &       0.0000  &  0.0000  &  0.0000   & 0.0000       	    \\

PPR &0.0417  &  0.0387  &  0.0553   & 0.0743       \vspace{1cm}   	  \\

 &  \multicolumn{4}{c} {Centrally symmetric Cauchy kernel $\&$ Shift skewing}  \\[.5mm]

Test    &     $\deltab=(0,0)$    &    $\deltab=(0,0.04)$     &     $\deltab=(0,0.08)$  & $\deltab=(0,0.12)$      \\
\hline
DepthRunH & 0.0410   & 0.1090   & 0.3817   & 0.7003    	   \\
DepthRunS & 0.0363  &  0.1087   & 0.3737   & 0.7120      	  \\
DepthRunSV &0.0370  &  0.1223  &  0.3750   & 0.7043     	   \\
BarS & 1.0000  &  1.0000   & 1.0000  &  1.0000            		    \\
Cassart&       0.0480  &  0.0557   & 0.0473  &  0.0463              	   \\

PPG &       0.0007  &  0.0023  &  0.0003   & 0.0013       	    \\

PPR &0.0333  &  0.0590  &  0.1750   & 0.4313          	 \vspace{1cm} \\

  \end{tabular}
  }
  \caption{Numerical rejection frequencies (out of 3,000 replications) under Azzalini-skewed spherically symmetric and shifted centrally symmetric bivariate Cauchy distributions, of the proposed depth-based runs tests, the \cite{Bar1991}-type spherical test, the pseudo-Gaussian test from \cite{Cas2007}, and the projection pursuit tests from \cite{Blo1989}, for samples of size~$n=100$.}
\label{Tsimu1}
\end{table}


\clearpage

\begin{thebibliography}{1}
\vspace{2mm}



\bibitem{AzzCap2003}
\mbox{A.} Azzalini $\&$ \mbox{A.} Capitanio (2003). Distributions generated by perturbation of symmetry with emphasis on a multivariate skew-$t$ distribution. \emph{Journal of the Royal Statistical Society Series B}, {\bf 65}, 367--389.

\bibitem{AzzDal1996}
\mbox{A.} Azzalini $\&$ \mbox{A.} Dalla Valle (1996). The multivariate skew-normal distribution. \emph{Biometrika}, {\bf 83}, 715--726.

\bibitem{Bar1991}
\mbox{L.} Baringhaus (1991). Testing for spherical symmetry of a multivariate distribution. \emph{The Annals of Statistics}, {\bf 19}, 899--917. 

\bibitem{Blo1989}
\mbox{D.K.} Blough (1989). Multivariate symmetry via projection pursuit. \emph{Annals of the Institute of Statistical Mathematics}, {\bf 41}, 461--475.

\bibitem{Cas2007}
\mbox{D.} Cassart (2007). Optimal tests for symmetry. Unpublished thesis. Univ. libre de Bruxelles, Brussels.

\bibitem{CohMen1988}
\mbox{J.P.} Cohen $\&$ \mbox{S.S.} Menjoge (1988). One-sample run tests of symmetry. \emph{Journal of Statistical Planning and Inference}, {\bf 18}, 93--100.

\bibitem{GhoRuy1992}
\mbox{S.} Ghosh $\&$ \mbox{F.H.} Ruymgaart (1992). Applications of empirical characteristic functions in some multivariate problems. \emph{The Canadian Journal of Statistics}, {\bf 20}, 429--440.

\bibitem{Heaetal1995}
\mbox{C.R.} Heathcote, \mbox{S.T.} Rachev $\&$ \mbox{B.} Cheng (1995). Testing multivariate symmetry. \emph{Journal of Multivariate Analysis}, {\bf 54}, 91--112.

\bibitem{Hen1993}
\mbox{N.} Henze (1993). On the consistency of a test for symmetry based on a runs statistic. \emph{Journal of Nonparametric Statistics}, {\bf 3}, 195--199.

\bibitem{Henetal2003}
\mbox{N.} Henze, \mbox{B.} Klar $\&$ \mbox{S.G.} Meintanis (2003). Invariant tests for symmetry about an unspecified point based on the empirical characteristic function. \emph{Journal of Multivariate Analysis}, {\bf 87}, 275--297.

\bibitem{JonPew2009}
\mbox{M.C.} Jones $\&$ \mbox{A.} Pewsey (2009). Sinh-arcsinh distributions. \emph{Biometrika}, {\bf 96}, 761--780.


\bibitem{Liu1990}
\mbox{R.Y.} Liu (1990). On a notion of data depth based on random simplices. \emph{The Annals of Statistics}, {\bf 18}, 405--414.

\bibitem{LiuSin1993}
\mbox{R.Y.} Liu, $\&$ \mbox{K.} Singh (1993). A quality index based on data depth and multivariate rank tests. \emph{Journal of the American Statistical Association}, {\bf 88}, 252--260.

\bibitem{Mar1999}
\mbox{J.} Marden (1999). Multivariate rank tests. \emph{In Multivariate Analysis, Design of Experiments, and Survey Sampling, \mbox{ed. S.} Ghosh, New York: Marcel Dekker}, 401--432.

\bibitem{McW1990}
\mbox{T.P.} WcWilliams (1990). A distribution-free test for symmetry based on a runs statistic. \emph{Journal of the American Statistical Association}, {\bf 85}, 1130--1133.

\bibitem{ModGas1996}
\mbox{R.} Modarres $\&$ \mbox{J.L.} Gastwirth (1996). A modified runs test for symmetry. \emph{Statistics and Probability Letters}, {\bf 31}, 107--112.

\bibitem{NeuZhu1998}
\mbox{G.} Neuhaus $\&$ \mbox{L.-X.} Zhu  (1998). Permutation tests for reflected symmetry. \emph{Journal of Multivariate Analysis}, {\bf 67}, 129--153.

\bibitem{Neu2013}
\mbox{M.} Neumann (2013). A central limit theorem for triangular arrays of weakly dependent random variables, with applications in statistics. \emph{ESAIM: Probability and Statistics}, {\bf 17}, 120-134.

\bibitem{Oja1983}
\mbox{H.} Oja (1983). Descriptive statistics for multivariate distributions. \emph{Statistics and Probability Letters}, {\bf 1}, 327--332.


\bibitem{PaiVan2013}
\mbox{D.} Paindaveine $\&$ \mbox{G.} Van Bever (2013). From depth to local depth : a focus on centrality. \emph{Journal of the American Statistical Association}, {\bf 105}, 1105--1119.

\bibitem{Ser2006B}
\mbox{R.J.} Serfling (2006). Multivariate symmetry and asymmetry. \emph{In Encyclopedia of Statistical Sciences, Second Edition (\mbox{S.} Kotz, \mbox{N.} Balakrishnan, \mbox{C.B.} Read and \mbox{B.} Vidakovic, eds.), Vol.~8, Wiley}, 5338--5345.

\bibitem{Tuk1975}
\mbox{J.} Tukey (1975). Mathematics and picturing data. \emph{In Proceedings of the 1975 International Congress of Mathematics 2}, 523--531.

\bibitem{Tyl1987}
\mbox{D.E.} Tyler (1987). A distribution-free $M$-estimator of multivariate scatter. \emph{The Annals of Statistics}, {\bf 15}, 234--251.



\bibitem{Zuo2003}
\mbox{Y.} Zuo (2003). Projection-based depth functions and associated medians. \emph{The Annals of Statistics}, {\bf 31}, 1460--1490.

\bibitem{ZuoSer2000A}
\mbox{Y.} Zuo $\&$ \mbox{R.J.} Serfling (2000a). General notions of statistical depth function. \emph{The Annals of Statistics}, {\bf 28}, 461--482.

\bibitem{ZuoSer2000B}
\mbox{Y.} Zuo $\&$ \mbox{R.J.} Serfling (2000b). Structural properties and convergence results for contours of sample statistical depth functions. \emph{The Annals of Statistics}, {\bf 28}, 483--499.


\end{thebibliography}
\end{document}